\documentclass[aps, prd, floatfix, nofootinbib, superscriptaddress, twocolumn]{revtex4-1}

\usepackage{latexsym}
\usepackage{amsmath}
\usepackage{amssymb}
\usepackage{amsfonts}

\usepackage[mathscr,scaled=1.15]{urwchancal}
\DeclareFontFamily{OT1}{pzc}{}
\DeclareFontShape{OT1}{pzc}{m}{it}%
{<-> s * [1.15] pzcmi7t}{}
\DeclareMathAlphabet{\mathpzc}{OT1}{pzc}{m}{it}

\usepackage{color}

\usepackage{supertabular}
\usepackage{placeins}
\usepackage{epsfig}
\usepackage{graphicx}

\definecolor{purple}{rgb}{0.5,0,0.5}
\definecolor{blue}{rgb}{0.0,0,0.9}
\definecolor{prdblue}{rgb}{0.133,0.118,0.498}
\usepackage[colorlinks=true, pdfstartview=FitV, linkcolor=prdblue, citecolor= prdblue, urlcolor=prdblue]{hyperref}

\begin{document}

% Use the \preprint command to place your local institutional report number in the upper righthand corner of the title
% page in preprint mode. Multiple \preprint commands are allowed. Use the 'preprintnumbers' class option to override
% journal defaults to display numbers if necessary
%\preprint{}

%\title{Symmetry-preserving calculation of the pion's valence-quark distribution}
%\title{Unification of continuum and lattice predictions of the pion's valence distribution}
\title{$\,$\\[-7ex]\hspace*{\fill}{\normalsize{\sf\emph{Preprint no}. NJU-INP 003/19}}\\[1ex]
Symmetry, symmetry breaking, and pion parton distributions}

\author{Minghui Ding}
\affiliation{School of Physics, Nankai University, Tianjin 300071, China}
\affiliation{European Centre for Theoretical Studies in Nuclear Physics
and Related Areas (ECT$^\ast$) and Fondazione Bruno Kessler\\ Villa Tambosi, Strada delle Tabarelle 286, I-38123 Villazzano (TN) Italy}

\author{Kh\'epani Raya}
\affiliation{School of Physics, Nankai University, Tianjin 300071, China}

\author{Daniele Binosi}
\affiliation{European Centre for Theoretical Studies in Nuclear Physics
and Related Areas (ECT$^\ast$) and Fondazione Bruno Kessler\\ Villa Tambosi, Strada delle Tabarelle 286, I-38123 Villazzano (TN) Italy}

\author{Lei Chang}
\email[]{leichang@nankai.edu.cn}
\affiliation{School of Physics, Nankai University, Tianjin 300071, China}

\author{Craig D.~Roberts}
\email[]{cdroberts@nju.edu.cn}
\affiliation{School of Physics, Nanjing University, Nanjing, Jiangsu 210093, China}
\affiliation{Institute for Nonperturbative Physics, Nanjing University, Nanjing, Jiangsu 210093, China}

\author{Sebastian M. Schmidt}
%\email[]{s.schmidt@fz-juelich.de}
\affiliation{
Institute for Advanced Simulation, Forschungszentrum J\"ulich and JARA, D-52425 J\"ulich, Germany}


\date{12 February 2020}
%\date{15 January 2020}
%\date{13 May 2019}
%\date{07 April 2019}

\begin{abstract}
%A symmetry-preserving approach to the two valence-body continuum bound-state problem is used to calculate the valence, glue and sea distributions within the pion; unifying them with, \emph{inter alia}, the electromagnetic pion elastic and transition form factors.  %
%The analysis reveals the following light-front momentum fractions at the scale $\zeta=2\,$GeV:
%$\langle x_{\rm valence} \rangle = 0.48(3)$,
%$\langle x_{\rm glue} \rangle  = 0.41(2)$,
%$\langle x_{\rm sea} \rangle  = 0.11(2)$;
%and despite hardening induced by the emergent phenomenon of dynamical chiral symmetry breaking, the valence-quark distribution function, ${\mathpzc q}^\pi(x)$, exhibits the $x\simeq 1$ behaviour predicted by quantum chromodynamics (QCD).  After evolution to $\zeta=5.2\,$GeV, the prediction for ${\mathpzc q}^\pi(x)$ matches that obtained using lattice-regularised QCD; hence two disparate treatments are now seen to yield the same prediction.  This confluence should both stimulate improved analyses of existing data and aid in motivating and supporting efforts to obtain new data on the pion distribution functions at existing and anticipated facilities.
%%%%
Pion valence, glue and sea distributions are calculated using a continuum approach to the two valence-body bound-state problem.  Since the framework is symmetry preserving, physical features of the distributions are properly expressed.  The analysis reveals that the emergent phenomenon of dynamical chiral symmetry breaking causes a hardening of the valence-quark distribution function, ${\mathpzc q}^\pi(x)$.  Nevertheless, this distribution exhibits the $x\simeq 1$ behaviour predicted by quantum chromodynamics (QCD).
At the scale $\zeta_2:=2\,$GeV, the following momentum fractions are predicted:
$\langle x_{\rm valence} \rangle = 0.48(3)$,
$\langle x_{\rm glue} \rangle  = 0.41(2)$,
$\langle x_{\rm sea} \rangle  = 0.11(2)$.
Evolving to $\zeta=5.2\,$GeV, the result for ${\mathpzc q}^\pi(x)$ agrees with that computed using lattice QCD.  These outcomes should both spur improved analyses of existing experiments and stimulate efforts to obtain new data on the pion distribution functions using available and envisioned facilities.
\end{abstract}

\maketitle

%%%%%%%%%%%%%%%%%%%%%%%%%%%%%%%%%%%%%%%%%%%%%%%%%%%%%%%%%%%%%%%%%%%%%%%%%%%%%%%%%%%%%%%%%%%%%%%%%%%%%%%%%%%%%%%%%%%%%%%
% 4500 words

%\noindent\textbf{1.$\;$Introduction}.
\section{Introduction}
\label{Introduction}
Simply regarding their valence-quark content, pions are Nature's simplest hadrons:
\begin{equation}
\pi^+  \sim u\bar d\,,\;
\pi^- \sim d \bar u\,, \;
\pi^0 \sim u\bar u - d\bar d\,;
\end{equation}
but this appearance is misleading.  Despite being hadrons, their physical masses are similar to that of the $\mu$-lepton; and the pion masses vanish in the chiral limit, \emph{i.e}.\ in the absence of a Higgs coupling for $u$- and $d$-quarks: they are the Nambu-Goldstone (NG) modes generated by dynamical chiral symmetry breaking (DCSB) in the Standard Model.  This dichotomous character -- simultaneous existence as NG-bosons and bound-states -- entails that the challenges of charting and explaining pion structure are of central importance in modern physics.  These problems are made more difficult by the crucial role of symmetries and their breaking patterns in determining pion properties, which must be properly incorporated and veraciously expressed in any theoretical treatment.

Given the pions' simple valence-quark content, a basic quantity in any discussion of their structure is the associated distribution function, ${\mathpzc q}^\pi(x;\zeta)$.  This density charts the probability that a valence ${\mathpzc q}$-quark in the pion carries a light-front fraction $x$ of the system's total momentum; and one of the earliest predictions of the parton model, augmented by features of perturbative quantum chromodynamics (pQCD), is \cite{Ezawa:1974wm, Farrar:1975yb, Berger:1979du}:
\begin{equation}
\label{PDFQCD}
{\mathpzc q}^{\pi}(x;\zeta =\zeta_H) \sim (1-x)^{2}\,,
\end{equation}
where $\zeta_H$ is an energy scale characteristic of nonperturbative dynamics.  Moreover, the exponent evolves as $\zeta$ increases beyond $\zeta_H$, becoming $2+\gamma$, where $\gamma\gtrsim 0$ is an anomalous dimension that increases logarithmically with $\zeta$.  (In the limit of exact ${\mathpzc G}$-parity symmetry, which is a good approximation in the Standard Model, $u^{\pi^+}(x) = \bar d^{\pi^+}(x)$, etc. Hence it is only necessary to discuss one unique distribution.)

Owing to the validity of factorisation in QCD, ${\mathpzc q}^\pi(x)$ is measurable in $\pi N$ Drell-Yan experiments \cite{Badier:1980jq, Badier:1983mj, Betev:1985pg, Falciano:1986wk, Guanziroli:1987rp, Conway:1989fs, Heinrich:1991zm}. However, conclusions drawn from analyses of these experiments have proved controversial \cite{Holt:2010vj}.
For instance, using a leading-order (LO) pQCD analysis of their data, Ref.\,\cite{Conway:1989fs} (the E615 experiment) reported ($\zeta_5 = 5.2\,$GeV):
\begin{align}
{\mathpzc q}_{\rm E615}^{\pi}(x; \zeta_5) &\sim (1-x)^{1}\,,
\end{align}
a marked contradiction of Eq.\,\eqref{PDFQCD}.  Subsequent calculations \cite{Hecht:2000xa} confirmed Eq.\,\eqref{PDFQCD} and eventually prompted reconsideration of the E615 analysis, with the result that, at next-to-leading order (NLO) and including soft-gluon resummation \cite{Wijesooriya:2005ir, Aicher:2010cb}, the E615 data can be viewed as being consistent with Eq.\,\eqref{PDFQCD}.

Notwithstanding these advances, uncertainty over Eq.\,\eqref{PDFQCD} will remain until other analyses of the E615 data incorporate threshold resummation effects and, crucially, new data are obtained. Prospects for the latter are good because relevant tagged deep-inelastic scattering experiments are approved at the Thomas Jefferson National Accelerator Facility \cite{Keppel:2015, Keppel:2015B, McKenney:2015xis} and the goal has high priority at other existing and anticipated facilities \cite{Petrov:2011pg, Peng:2016ebs, Peng:2017ddf, Horn:2018fqr, Denisov:2018unj}.

Meanwhile, progress in theory continues.  Novel perspectives and algorithms within lattice-regularised QCD (lQCD) \cite{Liu:1993cv, Ji:2013dva, Radyushkin:2016hsy, Radyushkin:2017cyf, Chambers:2017dov} are beginning to yield results for the pointwise behaviour of the pion's valence-quark distribution \cite{Xu:2018eii, Chen:2018fwa, Karthik:2018wmj, Sufian:2019bol, Joo:2019bzr}, offering promise for information beyond the lowest few moments \cite{Best:1997qp, Detmold:2003tm, Brommel:2006zz, Oehm:2018jvm}.

Extensions of the continuum analysis in Ref.\,\cite{Hecht:2000xa} are also yielding new insights.  For example:
a class of corrections to the handbag-diagram representation of the virtual-photon--pion forward Compton scattering amplitude has been identified and shown to restore basic symmetries in calculations of ${\mathpzc q}^\pi(x;\zeta)$ \cite{Chang:2014lva};
and the corrected expression has been used to compute all valence-quark distribution functions in the pion and kaon \cite{Chen:2016sno}, with the results indicating that the gluon content of the pion is significantly greater than that of the kaon.

This last feature owes to the mechanism responsible for the emergence of mass in the Standard Model.  Studies of meson properties \cite{Ding:2015rkn, Gao:2017mmp, Chen:2018rwz, Ding:2018xwy} indicate that the $s$-quark defines a boundary: emergent mass generation dominates for $\hat m < \hat m_s$, but the Higgs-mass prevails on $\hat m \gtrsim \hat m_s$, where $\hat m$ is the renormalisation group invariant current-mass for a given quark.  Hence, comparisons between the properties of systems containing only light quarks and those with one (or more) $s$-quark(s) are well suited to exposing effects of dynamical mass generation.

Given its potential for validating such observations, there is renewed interest in measuring $u^K(x)/u^\pi(x)$ \cite{Keppel:2015, Keppel:2015B, Peng:2016ebs, Peng:2017ddf, Horn:2018fqr, Denisov:2018unj}.  Only one data set currently exists \cite{Badier:1980jq}.  It is old (from 1980) and limited; hence, needs modernising and expanding in order to be effective in this new role.
%
%These observations have spurred new interest in measurements that can yield $u^K(x)/u^\pi(x)$ \cite{Keppel:2015, Keppel:2015B, Peng:2016ebs, Peng:2017ddf, Horn:2018fqr, Denisov:2018unj}, for which there is only one existing data set, from 1980 \cite{Badier:1980jq}.

The theory predictions also need updating, \emph{e.g}.\ the continuum results in Refs.\,\cite{Chang:2014lva, Chen:2016sno} were obtained using algebraic models for the elements needed to describe the Compton amplitude, \emph{i.e}.\ dressed-quark propagators, meson Bethe-Salpeter amplitudes and dressed-photon-quark vertex.  Therefore, following the recent theory developments, especially concerning the pion, herein we expand upon the calculation of pion parton distributions reported in Ref.\,\cite{Ding:2019qlr}, providing extensive explanations and including additional material that should, \emph{inter alia}, prove valuable in illuminating the formulation, analyses and results.  Notably, this approach to the two-body bound-state problem has successfully unified the treatment of the charged-pion-elastic and neutral-pion-transition form factors \cite{Chang:2013nia, Raya:2015gva, Raya:2016yuj, Gao:2017mmp, Ding:2018xwy}.  It has also been used to correlate continuum and lattice predictions for the electromagnetic form factors of charged pion-like mesons, thereby enabling an extrapolation of the lQCD results to the physical pion mass \cite{Chen:2018rwz}.

Our manuscript is arranged as follows.
Section~\ref{secFormal} describes the connection between pion Compton scattering and ${\mathpzc q}^\pi(x)$;
and Sec.\,\ref{secSymmetries} recapitulates the analysis of Ref.\,\cite{Chang:2014lva}, explaining the flaws of the handbag diagram as a tool for calculating valence-quark distributions and illuminating the corrections that repair its deficiencies and thus produce a symmetry-preserving approximation.
%We address this issue herein by first correcting a commonly used expression for the valence-quark distribution function and then illustrating its consequences with an algebraic model that incorporates salient features of QCD.
%
%extend Chang:2014lva ... same interaction that predicted pion form factor and unified lQCD calculations thereof to provide an up-to-date and most realistic prediction for the pion pdf, which properly respects all relevant symmetries.
%
Section~\ref{predictqpion} reports our calculation of ${\mathpzc q}^\pi(x)$ at $\zeta_H$: detailing the kernel used to solve the continuum bound-state problem; computation of the lowest six independent Mellin-moments; and reconstruction of ${\mathpzc q}^\pi(x)$ therefrom.  It also explains that, through a connection between the saturation value of QCD's process-independent effective charge and the one-loop running coupling, the hadronic scale is determined: $\zeta_H =0.30\,$GeV.
Evolution of ${\mathpzc q}^\pi(x;\zeta_H)$ to $\zeta/{\rm GeV}=2, 5.2$ for comparison with data, phenomenology and theory is described in Sec.\,\ref{secEvolution}; and predictions for the glue and sea momentum-distributions are obtained analogously using the singlet evolution equations.
Section~\ref{epilogue} provides a summary and offers perspectives.

%\pagebreak

\section{Quark distribution function}
\label{secFormal}
The hadronic tensor relevant to inclusive deep inelastic lepton-pion scattering may be expressed via two invariant structure functions \cite{Jaffe:1985je}.  With the incoming photon possessing momentum $q$ and the target pion, momentum $P$, then in the deep-inelastic (Bjorken) limit \cite{Bjorken:1968dy}, \emph{viz}.\
%$q^2\to\infty$, $P\cdot q \to -\infty$ but $x:= - q^2/[2 P \cdot q]$ fixed,
\begin{equation}
\label{BJlim}
q^2\to\infty\,,\; P\cdot q \to -\infty, \;\mbox{but}\;
x:= - q^2/[2 P \cdot q] \; %\frac{q^2}{2 P\cdot q}\;\;
\mbox{fixed}\,,
\end{equation}
that tensor is $(t_{\mu\nu} = \delta_{\mu\nu}-q_\mu q_\nu/q^2, P_\mu^{\,t}= t_{\mu\nu}P_\nu)$:
\begin{subequations}
\begin{align}
W_{\mu\nu}(q;P) & = F_1(x)\, t_{\mu\nu} - \frac{F_2(x)}{P\cdot q}
\,P_\mu^{\,t} P_\nu^{\,t}\,,\\
& F_2(x)  = 2 x F_1(x)\,.
\end{align}
\end{subequations}

$F_1(x)$ is the pion structure function, which provides access to the pion's quark distribution functions:
\begin{equation}
\label{qPDF}
F_1(x) = \sum_{{\mathpzc q}\in \pi} \, e_{\mathpzc q}^2 \, {\mathpzc q}^\pi(x)\,,
\end{equation}
where $e_{\mathpzc q}$ is the quark's electric charge.  The sum in Eq.\,\eqref{qPDF} runs over all quark flavours; but in the $\pi^+$ it is naturally dominated by $u(x)$, $\bar d(x)$.  Moreover, in the $\mathpzc{G}$-parity symmetric limit, which we employ throughout, $u(x)=\bar d(x)$.  (Bjorken-$x$ is equivalent to the light-front momentum fraction of the struck parton.)  Using the optical theorem, the structure function is given by the imaginary part of the virtual-photon--pion forward Compton scattering amplitude: $\gamma^\ast(q) + \pi(P) \to \gamma^\ast(q) + \pi(P)$.

\section{Symmetries and the pion valence-quark distribution function}
\label{secSymmetries}
\subsection{Charged-pion form factor}
In order to elucidate the role of symmetries in developing an approximation to the $\gamma^\ast \pi$ forward Compton amplitude, which is the basis for any computation of ${\mathpzc q}^\pi(x)$, we first consider the simpler problem of the pion electromagnetic form factor, $F_\pi(Q^2)$.  In both cases, one aims to expose structural features of a system characterised by two valence-parton degrees-of-freedom.  A useful framework for studying such problems in quantum field theory is provided by the Dyson-Schwinger equations (DSEs) \cite{Roberts:1994dr}, with the one-body gap equation and two-body Bethe-Salpeter equation (BSE) playing leading roles.

The DSEs are a collection of coupled equations and a tractable problem is obtained once a truncation scheme is specified.  A weak-coupling expansion reproduces perturbation theory; but although valuable in the analysis of large momentum transfer phenomena in QCD, it cannot be used to obtain nonperturbative information.  A symmetry-preserving scheme applicable to hadrons was introduced in Refs.\,\cite{Munczek:1994zz, Bender:1996bb} and has subsequently been exploited \cite{Roberts:2000aa, Maris:2003vk, Horn:2016rip, Eichmann:2016yit} and refined \cite{Chang:2009zb, Fischer:2009jm, Chang:2011ei, Binosi:2014aea, Binosi:2016wcx, Williams:2015cvx, Binosi:2016rxz}.  The basic point is that the Bethe-Salpeter kernel appropriate to a given two-valence-body problem is computable once that of the one-body gap equation is specified.  Following these procedures, one guarantees,  \emph{inter} \emph{alia}, that all Ward-Green-Takahashi (WGT) identities \cite{Ward:1950xp, Green:1953te, Takahashi:1957xn, Takahashi:1985yz} are preserved, without fine-tuning, and thereby ensures, \emph{e.g}.\ current-conservation and the appearance of NG modes in connection with DCSB.  These qualities are essential in connection with studies of electromagnetic interactions involving pions (and other pseudoscalar mesons).  (\emph{N.B}.\ Calculations, such as ours, are typically performed using the Poincar\'e-covariant Landau gauge because, \emph{inter alia}, this ensures that preservation of symmetries is readily tracked and elucidated via the WGT identities.)

\begin{figure}[t]
\centerline{%
\includegraphics[clip, width=0.325\textwidth]{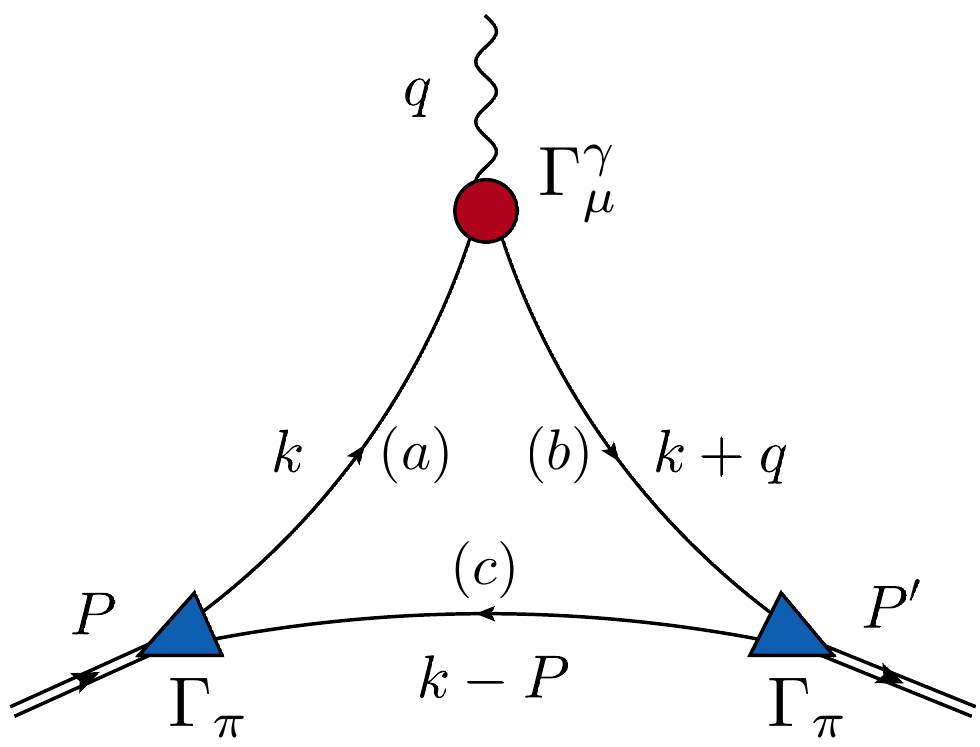}}
\caption{\label{RLFpiQ2}  RL truncation for the charged-pion electromagnetic form factor, Eq.\,\eqref{RLFpi}: triangles (blue) -- pion Bethe-Salpeter amplitudes; circle (red) -- amputated dressed-photon-quark vertex; and interior lines -- dressed-quark propagators.  Poincar\'e-covariance and electromagnetic current conservation, \emph{inter alia}, are guaranteed so long as each of these elements is computed in RL truncation.  For later use, we define line (\emph{a}) to be that carrying momentum $k$; line (\emph{b}), $k+q$; and line (\emph{c}), $k-P$.}
\end{figure}

The leading-order term in the procedure of Refs.\,\cite{Munczek:1994zz, Bender:1996bb} is the rainbow-ladder (RL) truncation.  Widely used, it is accurate for an array of systems and properties; in particular, those of ground-state flavour-nonsinglet pseudoscalar mesons because corrections in these channels largely cancel owing to the parameter-free preservation of relevant WGT identities ensured by this scheme.

Even before it was recognised as part of a systematic procedure, RL truncation was used as the basis for a calculation of $F_\pi$ \cite{Roberts:1994hh}.  As argued therein, to obtain the form factor at this level in the symmetry-preserving truncation, one computes the matrix element depicted in Fig.\,\ref{RLFpiQ2}:
\begin{eqnarray}
\nonumber
K_\mu F_{\pi}(Q^2) & = & N_c {\rm tr} %\int_{dk}^\Lambda
\int_{dk}\,
i\chi_\mu(k+q, k) \\
&& \times i\Gamma_{\pi}(k_i;P)\,S(k-P)\,i\Gamma_{\pi}(k_f;-P^\prime)\,, \quad\label{RLFpi}
\end{eqnarray}
where $q=P^\prime-P$ is the incoming photon momentum ($Q^2=q^2$), $2 K = P^\prime + P$;
%$p_i = P$, $p_f = P+q =:P^\prime$,
$P^2=-m_\pi^2=(P^\prime)^2$;
$k_i= k - P/2$, $k_f = k-P/2+q/2$;
$N_c=3$;
the trace is over spinor indices;
and $\int_{dk} := \int \frac{d^4 k}{(2\pi)^4}$ is a translationally invariant regularisation of the integral.
%
%(As noted above, we work in the ${\mathpzc G}$-parity symmetry limit.)

In Eq.\,\eqref{RLFpi},
$S$ is the $u=d$ dressed-quark propagator, computed in rainbow truncation;
$\Gamma_\pi$ is the pion Bethe-Salpeter amplitude, computed with a rainbow-ladder kernel;
and $\chi_\mu(k+q,k)=S(k+q)\Gamma_\mu(k+q,k) S(k)$, with $\Gamma_\mu$ the amputated dressed-photon-quark vertex, computed using the same kernel.

In RL truncation, there are no corrections to Eq.\,\eqref{RLFpi}.
To see this, suppose one were to add a gluon emitted from line (\emph{a}) and reabsorbed by line (\emph{a}).  This would be over-counting because the contribution is already included in the rainbow truncation computation of the dressed-quark propagator.
Suppose next that a gluon is emitted by line (\emph{a}) and absorbed by line (\emph{c}).  That would also be over-counting because such a contribution is already contained in the RL-truncation result for $\Gamma_\pi$.
Indeed, no matter which line or lines one chooses to emit and reabsorb a single gluon, the contribution generated is already included in $S$, $\Gamma_\pi$ or $\Gamma_\mu$.
Consequently, Fig.\,\ref{RLFpiQ2} depicts the complete RL result for $F_\pi(Q^2)$.  It is the basis for a calculation of this form factor on the entire domain of spacelike $Q^2$ \cite{Chang:2013nia, Gao:2017mmp, Chen:2018rwz}, which agrees with existing data \cite{Horn:2006tm, Horn:2007ug, Huber:2008id, Blok:2008jy} and predicts that QCD scaling violations will be seen in experiments that reach $Q^2 \gtrsim 8\,$GeV$^2$.

\begin{figure*}[t]
\includegraphics[clip, width=0.325\textwidth]{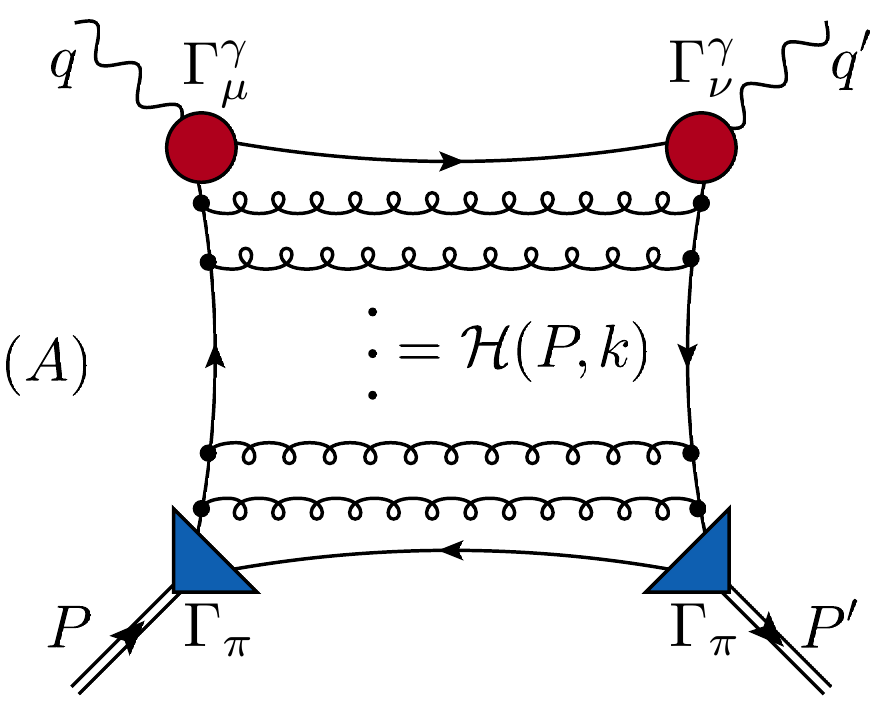}
\includegraphics[clip, width=0.325\textwidth]{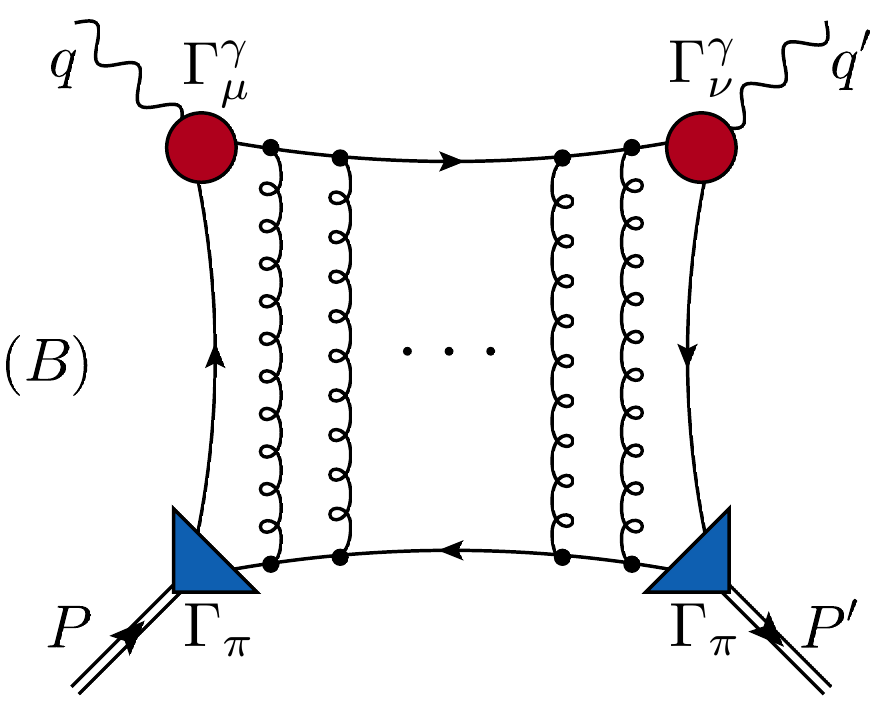}
\includegraphics[clip, width=0.325\textwidth]{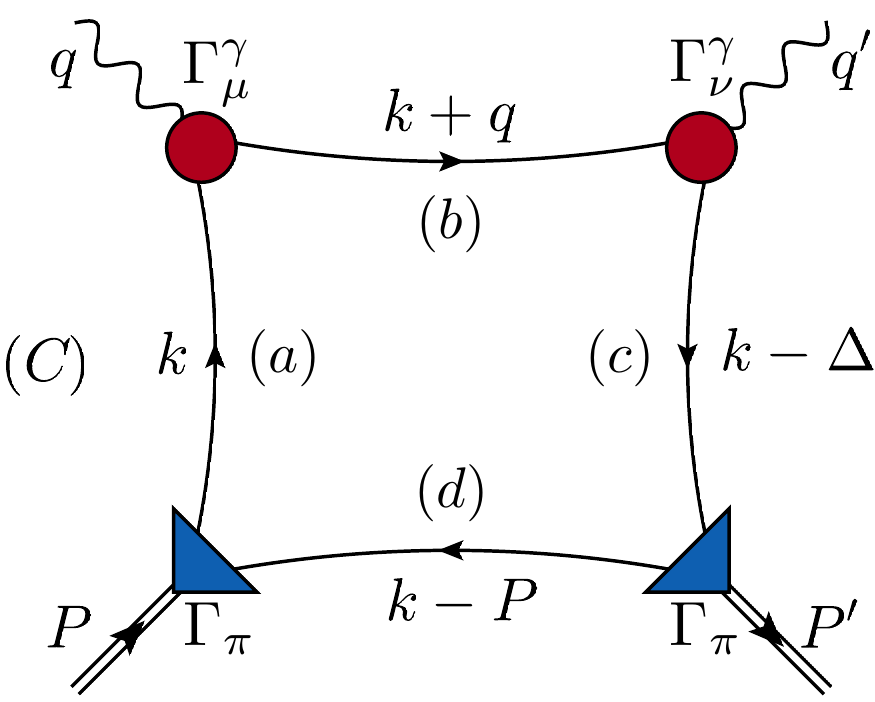}
\caption{\label{figCompton}
Collection of diagrams required to complete a sym\-metry-preserving RL calculation of pion Compton scattering.
Amplitude-One (\emph{S1}) = (\emph{A})+(\emph{B})-(\emph{C}).  The ``dots'' in (\emph{A}) and (\emph{B}) indicate summation of infinitely many ladder-like rungs, beginning with zero rungs.
The other two amplitudes are obtained as follows: (\emph{S2}) -- switch vertices to which $q$ and $q^\prime$ are attached; and (\emph{S3}) -- switch vertex insertions associated with $q^\prime$ and $P^\prime$.
In all panels: triangles (blue) -- pion Bethe-Salpeter amplitudes; circles (red) -- amputated dressed-photon-quark vertices; and interior lines -- dressed-quark propagators.  $\Delta = q^\prime - q$.
Poincar\'e-covariance and electromagnetic current conservation, \emph{inter alia}, are guaranteed so long as each of these elements is computed in RL truncation.
For later use, using (\emph{C}), we define line (\emph{a}) to be that carrying momentum $k$; line (\emph{b}), $k+q$; line (\emph{c}), $k-\Delta$; and line (\emph{d}), $k-P$.
}
\end{figure*}

\subsection{Pion valence-quark distribution function}
Viewed simply, RL truncation represents $F_\pi$ as a three-point function: there are three compound vertices in Fig.\,\ref{RLFpiQ2}.  Counting in the same way, the photon-pion Compton amplitude is a four-point function; and anyone desiring to supply predictions for ${\mathpzc q}^\pi(x)$ that are consistent with those for $F_\pi$ is immediately presented with the challenge of writing the complete RL truncation for this four-point function.  The solution to that problem was presented almost twenty years ago, in connection with the treatment of $\pi \pi \to \pi\pi$ \cite{Bicudo:2003fp, Bicudo:2001jq}.  Translated to the pion Compton amplitude \cite{Chang:2014lva}, the complete (symmetry-preserving) RL truncation is given by permutations of the three diagrams illustrated in Fig.\,\ref{figCompton}.  (An extension to nucleon Compton scattering is described elsewhere \cite{Eichmann:2012mp}.)

That the collection of diagrams in Fig.\,\ref{figCompton} is necessary and sufficient to generate the complete, symmetry-preserving RL treatment of $\gamma\pi \to\gamma\pi$ is readily made apparent.
For example, suppose one were to add a gluon emitted from line (\emph{a}) and reabsorbed by line (\emph{a}).  This would be over-counting because the contribution is already included in the rainbow truncation dressed-quark propagator.
Suppose next that an additional gluon is emitted by line (\emph{a}) and absorbed by line (\emph{d}).  That would also be over-counting because this diagram is already contained in the RL result for $\Gamma_\pi$.
Now imagine that a new gluon is emitted by line (\emph{a}) and absorbed by line (\emph{b}).  That is over-counting because such contributions are contained in the RL dressed-photon-quark vertex.
One must also consider a gluon emitted by line (\emph{a}) and absorbed by line (\emph{c}).  This is one of the summed diagrams represented by Fig.\,\ref{figCompton}(\emph{A}); and Fig.\ref{figCompton}(\emph{B}) represents the sum of contributions obtained by laddered gluons between lines (\emph{b}) and (\emph{d}) in Fig.\ref{figCompton}(\emph{C}).
Allowing only such one-gluon-like exchange effects, which is the basic feature of RL truncation, then there are no distinct additional contributions.
On the other hand, if any one of the contributions described and illustrated here is neglected in a given calculation, then that calculation explicitly breaks an array of relevant symmetries.

% it is evident that the virtual Compton amplitude in RL truncation should be built from permutations of the three diagrams illustrated in Fig.\,\ref{figCompton} \cite{Binosi:2003yf}.  This collection is necessary and sufficient to ensure preservation of the relevant WGT identities so long as the dressed-quark propagators, pion Bethe-Salpeter amplitudes and dressed--quark-photon vertices, appearing in the diagrams, are all computed in RL truncation.

Consider now the $\gamma^\ast \pi$ forward pion Compton amplitude in the Bjorken limit, Eq.\,\eqref{BJlim}.  The (\emph{S3}) permutation of the diagrams in Fig.\,\ref{figCompton} corresponds to a collection of so-called \emph{cat's ears} contributions.  They are greatly suppressed compared to the other two permutations in the Bjorken limit; hence may be neglected.  The (\emph{S2}) permutation corresponds simply to symmetrising the incoming and outgoing photons and so need not explicitly be considered further.  Consequently, in computing ${\mathpzc q}^\pi(x;\zeta_H)$, one may focus solely on those diagrams drawn explicitly in Fig.\,\ref{figCompton}; namely, in RL truncation \cite{Chang:2014lva}:
\begin{align}
%\nonumber & \mbox{\rm RL~truncation:} \\
& \gamma^\ast(q) + \pi(P)  \to \gamma^\ast(q) +\pi(P)  \stackrel{\mbox{\rm Fig.\,\ref{figCompton}}}{=}
(A) + (B) - (C)\,. \label{ComptonRL}
\end{align}

In the forward and Bjorken limits, Fig.\,\ref{figCompton}(\emph{C}) is the textbook \emph{handbag} contribution to $\gamma^\ast \pi$ Compton scattering.  It has often been used alone to estimate ${\mathpzc q}^\pi(x;\zeta_H)$.   (See, e.g.\ Refs.\,\cite{Shigetani:1993dx, Davidson:1994uv, Bentz:1999gx, Dorokhov:2000gu, Hecht:2000xa} and citations therein and thereof.)
If the pion's Bethe-Salpeter amplitude is assumed to be momentum-independent\footnote{This is the result obtained using an internally-consistent, symmetry-preserving treatment of a vector$\otimes$vector contact interaction (CI) \cite{GutierrezGuerrero:2010md, Roberts:2010rn}.} and a Poincar\'e-invariant regularisation of the loop-integral is employed, then Fig.\,\ref{figCompton}(\emph{C}) yields a result for ${\mathpzc q}^\pi(x;\zeta_H)$ that preserves both the baryon-number and momentum sum-rules; namely,
\begin{subequations}
\label{eqSumRules}
\begin{align}
\int_0^1dx\,{\mathpzc q}^\pi(x;\zeta_H) = 1\,,\; \label{BaryonSum}\\
\int_0^1dx\,x {\mathpzc q}^\pi(x;\zeta_H) = \frac{1}{2}\,.\; \label{MomSum}
\end{align}
\end{subequations}
(The right-hand-side of Eq.\,\eqref{BaryonSum} remains unity under QCD evolution -- DGLAP \cite{Dokshitzer:1977, Gribov:1972, Lipatov:1974qm, Altarelli:1977}.)
%% http://www.scholarpedia.org/article/QCD_evolution_equations_for_parton_densities Parisi article
%
In fact, one finds \cite{Dorokhov:2000gu}
\begin{equation}
\label{qpiCI}
{\mathpzc q}_{\rm CI}^\pi(x;\zeta_H) \approx \theta(x)\theta(1-x)\,,
\end{equation}
where $\theta(x)$ is the Heaviside step function.  Eq.\,\eqref{qpiCI} describes a structureless pion, in which a given valence-quark carries all light-front-fractions of the pion's total momentum with equal probability.

If the regularisation scheme for the loop in Fig.\,\ref{figCompton}(\emph{C}) introduces a mass-scale and/or the quark-antiquark interaction is momentum-dependent, then the result obtained violates one or both of the sum rules in Eq.\,\eqref{eqSumRules} \cite{Shigetani:1993dx, Hecht:2000xa}.  Consequently,  Fig.\,\ref{figCompton}(\emph{C}) alone is a poor approximation when realistic interactions are used.

Consider now Fig.\,\ref{figCompton}(\emph{A}), which can be written thus:
\begin{equation}
\label{qvHklx}
{\mathpzc q}_{A}^\pi(x;\zeta_H) = N_c{\rm tr} \! \int_{dk} \,
%\delta(n\cdot k_\eta - x n\cdot P) \,
\delta_n^{x}(k_\eta)\, n\cdot \gamma \,{\cal H}_\pi(P,k)\,,
\end{equation}
where
%$N_c=3$ and the trace is over spinor indices; $\int_{dk} := \int \frac{d^4 k}{(2\pi)^4}$ is a translationally invariant regularisation of the integral;
$\delta_n^{x}(k_\eta):= \delta(n\cdot k_\eta - x n\cdot P)$; $n$ is a light-like four-vector, $n^2=0$, with $n\cdot P = -m_\pi$ in the pion rest frame; and $k_\eta = k + \eta P$, $k_{\bar\eta} = k - (1-\eta) P$, $\eta\in [0,1]$.  Owing to Poincar\'e covariance, no observable can legitimately depend on $\eta$, \emph{i.e}.\ the definition of the relative momentum.

In RL truncation, as illustrated in Fig.\,\ref{figCompton}(\emph{A}), ${\cal H}_\pi(P,k)$ in Eq.\,\eqref{qvHklx} is an infinite sum of laddered gluon-rungs, beginning with zero rungs.  Hence, one may write \cite{Nguyen:2011jy}
\begin{align}
\nonumber
{\mathpzc q}_{A}^\pi&(x;\zeta_H) = N_c  {\rm tr} \! \int_{dk} \!
 i\Gamma_\pi(k_\eta,-P)\\
& \times \,S(k_\eta)\, i\Gamma^n(k;x;\zeta_H) \, S(k_\eta)\, i\Gamma_\pi(k_{\bar\eta},P)\, S(k_{\bar\eta})\,,
\label{Eucl_pdf_LR_Ward}
\end{align}
where $\Gamma^n(k;x;\zeta_H)$ is a generalisation of the quark-photon vertex, describing a dressed-quark scattering from a zero momentum photon and determined by a RL Bethe-Salpeter equation with inhomogeneity $n\cdot\gamma \,\delta_n^{x}(k_\eta)$.
% ($Z_2$ is the dressed-quark renormalisation constant.)
% It would seem to be necessary to compute this as Z_2(\zeta_H,\Lambda).

\begin{figure}[t]
\centerline{%
\includegraphics[clip, width=0.325\textwidth]{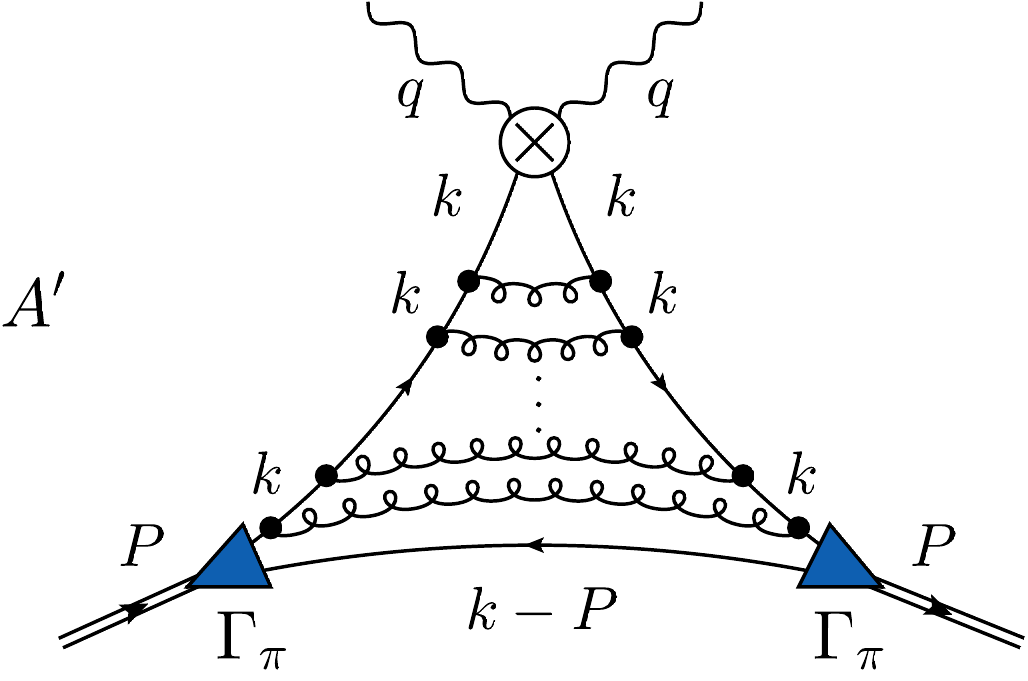}}
\smallskip

\centerline{%
\includegraphics[clip, width=0.325\textwidth]{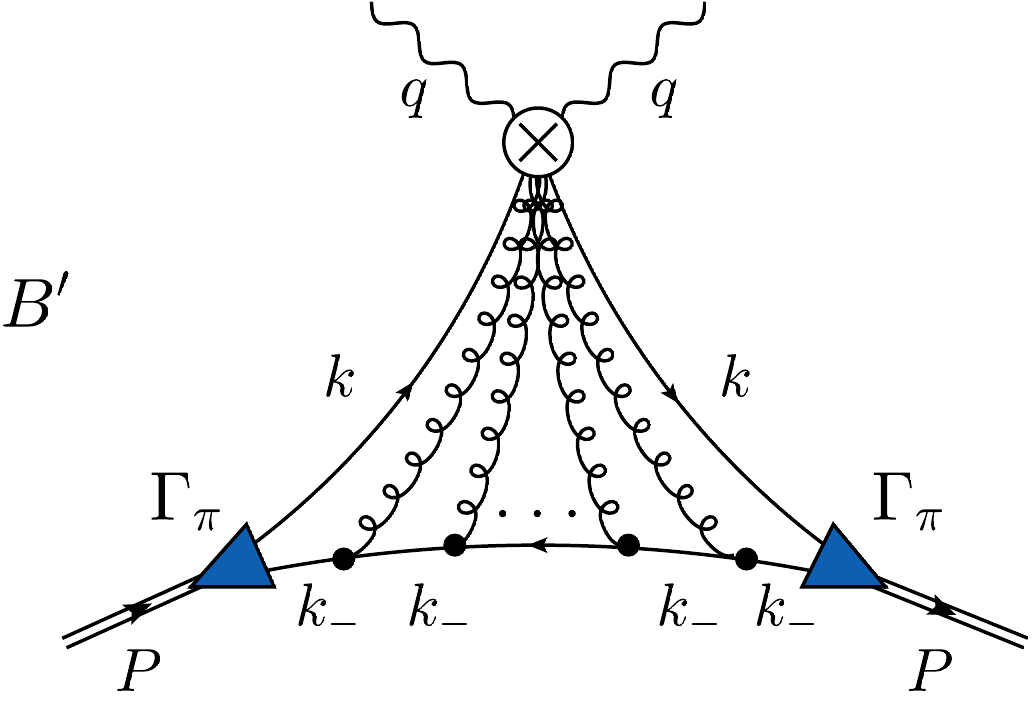}}
\caption{\label{Newqpi}
Employing the optical theorem, the diagrams in Fig.\,\ref{figCompton} yield these two contributions to ${\mathpzc q}^\pi(x)$: \emph{upper panel}, Eq.\,\eqref{Eucl_pdf_LR_Ward}; and \emph{lower panel}, Eq.\,\eqref{qBCPDF}.  The sum yields the completely symmetry-preserving RL truncation formula for ${\mathpzc q}^\pi(x)$.}
\end{figure}

Eq.\,\eqref{Eucl_pdf_LR_Ward} is depicted in Fig.\,\ref{Newqpi}($A^\prime$); and now a comparison with Fig.\,\ref{RLFpiQ2} makes manifest that the RL treatment of Fig.\,\ref{figCompton}(\emph{A}) is equivalent to the symmetry preserving analysis of the pion's electromagnetic form factor (at $Q^2=0$) \cite{Roberts:1994dr, Maris:2000sk}.
Furthermore, Eq.\,\eqref{Eucl_pdf_LR_Ward} ensures Eq.\,\eqref{BaryonSum} because
\begin{equation}
\int_0^1 dx \, \Gamma^n(k;x;\zeta_H) = n_\mu \Gamma_\mu(k,k) / n\cdot P\,;
\end{equation}
thus, using Eq.\,\eqref{RLFpi},
\begin{equation}
\label{qAunity}
\int_0^1dx\, {\mathpzc q}_A^\pi(x;\zeta_H) = F_\pi(Q^2=0)=1\,.
\end{equation}

On the other hand, as illustrated by existing calculations, \emph{e.g}.\ Refs.\,\cite{Hecht:2000xa, Nguyen:2011jy}, Eq.\,\eqref{Eucl_pdf_LR_Ward} violates Eq.\,\eqref{MomSum}.  Hence, as explained above, any result for ${\mathpzc q}^\pi(x;\zeta_H)$ obtained from Fig.\,\ref{figCompton}(\emph{A}) alone -- equivalently, Fig.\,\ref{Newqpi}($A^\prime$) -- is flawed because it violates basic symmetry constraints.
%\emph{i.e}.\ alone, as explained above, Fig.\,\ref{figCompton}(\emph{A}) -- equivalently, Fig.\,\ref{Newqpi}($A^\prime$) -- does \emph{not} define a complete symmetry-preserving RL treatment of $q^\pi(x)$.
Typical consequences include the following:
an overestimate of the sea and gluon content of a given bound-state;
erroneous estimates of the relative size of the valence-quark momentum fractions within different but related bound-states;
incorrect identification of $\zeta_H$, if this scale is used as a parameter to fit an empirically-determined distribution \cite{Jaffe:1980ti};
and since these errors are transmitted into the evolved distributions, a lack of credibility in any conclusions and interpretations drawn from the distributions.
Furthermore, the symmetry violations and associated errors are amplified by including the ${\cal H}(P,k)$ resummation in Fig.\,\ref{figCompton}(\emph{A}) [Fig.\,\ref{Newqpi}($A^\prime$)] alone because this unbalances the interferences that a fully-consistent RL truncation is guaranteed to preserve.
Consequently, less damage is done by working solely with Fig.\,\ref{figCompton}(\emph{C}).

We turn now to the contribution (\emph{B})-(\emph{C}) in Eq.\,\eqref{ComptonRL}, which has usually been overlooked in calculations of ${\mathpzc q}^\pi(x;\zeta_H)$; but whose importance was stressed and illustrated in Ref.\,\cite{Chang:2014lva}.
Given that the combination (\emph{B})-(\emph{C}) is crucial if the WGT identities are to be satisfied in a RL analysis of Compton scattering, let us consider their content.
A first observation is that (\emph{B$_0$})-(\emph{C})$\, =0$, \emph{i.e}.\ if one omits all terms from the ladder-like sum in Fig.\,\ref{figCompton}(\emph{B}) then it is completely cancelled by subtracting Fig.\,\ref{figCompton}(\emph{C}).  Hence, (\emph{B$_0$})-(\emph{C}) is a sum of infinitely many ladder-like rungs, beginning with one rung.
This is illustrated in Fig.\,\ref{Newqpi}$(B^\prime)$.
Studying this figure, the nature of (\emph{B})-(\emph{C}) becomes plain, \emph{viz}.\ it expresses the impact of the deep-inelastic event as felt by a dressed-quark line embedded \emph{within} the pion bound state.
Thinking perturbatively, one might imagine these processes to represent effects associated with initial/final-state interaction corrections to the handbag diagram and thus to be suppressed.  However, so long as the gluon exchanges are soft, which is the limit exposed by the optical theorem analysis, that is not the case because the resummation of ladder-like rungs is resonant.
Hence the contribution depicted in Fig.\,\ref{Newqpi}($B^\prime$) is of precisely the same order as that from Fig.\,\ref{Newqpi}($A^\prime$).
In fact, akin to the final state interactions that produce single spin asymmetries \cite{Brodsky:2002cx}, the $(B)$-$(C)$ contribution is leading-twist and its appearance and importance signal failure of the impulse approximation.

These considerations lead to the following form for the (\emph{B})-(\emph{C}) contribution to ${\mathpzc q}^\pi(x;\zeta_H)$ \cite{Chang:2014lva}:
\begin{align}
\nonumber {\mathpzc q}_{BC}^\pi(x;\zeta_H) & = N_c {\rm tr}\! \int_{dk}\! \Gamma_\pi^n(k_\eta,-P;\zeta_H)  \\
& \quad \times S(k_\eta)\Gamma_\pi(k_{\bar\eta},P)\, S(k_{\bar\eta})\,,
\label{qBCPDF}
\end{align}
where $\Gamma_\pi^n(k_\eta,-P;\zeta_H)$ is a ``pierced'' pion Bethe-Salpeter amplitude, computed by summing infinitely many insertions of $[\delta_n^{x}(k_\eta) n\cdot\partial_{k_\eta} S(k_\eta)]$, between sequentially-chosen adjacent gluon-rungs in the diagrammatic expansion of the pion amplitude.  Notably, independent of $\zeta_H$, as a consequence of symmetry preservation:
\begin{equation}
\int_0^1 dx\, {\mathpzc q}_{BC}^\pi(x;\zeta_H)  = 0\,.
\label{qDBzero}
\end{equation}

We can now write the complete expression for the pion valence-quark distribution function in RL truncation:
\begin{equation}
{\mathpzc q}^\pi(x;\zeta_H) = {\mathpzc q}_{\rm A}^\pi(x;\zeta_H) + {\mathpzc q}_{\rm BC}^\pi(x;\zeta_H) \,,
\end{equation}
\emph{i.e}.\ one sums the terms in Eqs.\,\eqref{Eucl_pdf_LR_Ward} and \eqref{qBCPDF}.
%This expression is nonzero in general.  It only vanishes when the pion's Bethe-Salpeter amplitude is independent of relative momentum; i.e., in the class of theories that employ a momentum-independent interaction, which includes models of the Nambu--Jona-Lasinio type \cite{Nambu:1961tp} and DSE-formulated analogues \cite{Roberts:2010rn}.

\section{Prediction for the pion valence-quark distribution function}
\label{predictqpion}
\subsection{Ward identity approximation for ${\mathpzc q}^\pi(x)$}
As illustrated in Ref.\,\cite{Bicudo:2001jq}, it is challenging to solve for the complete RL $u$ and $t$ channel scattering amplitudes depicted in Figs.\,\ref{figCompton}($A$), ($B$) and needed to describe $\gamma^\ast \pi \to \gamma^\ast \pi$.  Herein, we therefore use a simpler approach, employing the approximations introduced in Ref.\,\cite{Chang:2014lva}:
\begin{subequations}
\label{EqVtxModel}
\begin{align}
i\Gamma^n(k;x;\zeta_H) & = \delta_n^{x}(k_\eta) n\cdot\partial_{k_\eta} S^{-1}(k_\eta) \,,\\
\Gamma_\pi^n(k_\eta,-P;\zeta_H)  & = n\cdot\partial_{k_\eta} \Gamma_\pi(k_\eta,-P;\zeta_H)\,,
\end{align}
\end{subequations}
in which case
\begin{align}
& {\mathpzc q}^\pi(x;\zeta_H)  = N_c {\rm tr}\! \int_{dk}\! \delta_n^{x}(k_\eta) \nonumber \\
& \times \{n\cdot\partial_{k_\eta} \left[ \Gamma_\pi(k_\eta,-P) S(k_\eta) \right]\}
\Gamma_\pi(k_{\bar\eta},P)\, S(k_{\bar\eta})\,.
\label{qFULL}
\end{align}
Improvement upon Eqs.\,\eqref{EqVtxModel} will be canvassed in future, following Ref.\,\cite{Bicudo:2001jq}.  However, as shown therein, in Ref.\,\cite{Chang:2014lva}, and remarked above: the symmetry-preserving nature of our treatment, Fig.\,\ref{figCompton}, ensures cancellations between terms.  Hence, corrections in the $u$ channel largely compensate those at equal order in the $t$ channel thereby ensuring accuracy of Eq.\,\eqref{qFULL}.

It is straightforward to prove algebraically that the result obtained using Eq.\,\eqref{qFULL} is:
independent of $\eta$;
ensures
\begin{equation}
\label{xoneminusx}
{\mathpzc q}^\pi(x;\zeta_H) = {\mathpzc q}^\pi(1-x;\zeta_H)\,;
\end{equation}
satisfies Eqs.\,\eqref{eqSumRules};
and possesses defined subcomponents that comply with Eqs.\,\eqref{qAunity},  \eqref{qDBzero}.

\subsection{Computing the inputs for ${\mathpzc q}^\pi(x)$}
In order to calculate ${\mathpzc q}^\pi(x;\zeta_H)$ from Eq.\,\eqref{qFULL} one must know the dressed light-quark propagator and pion Bethe-Salpeter amplitude.  Algebraic \emph{Ans\"atze} were employed in Ref.\,\cite{Chang:2014lva}.  In contrast, herein we follow Ref.\,\cite{Chen:2018rwz} and use realistic numerical solutions.  Consequently, the result for ${\mathpzc q}^\pi(x;\zeta_H)$ is completely determined once an interaction kernel is specified for the RL Bethe-Salpeter equation.

We use the interaction explained in Ref.\,\cite{Qin:2011dd, Qin:2011xq}:
 \begin{subequations}
\label{KDinteraction}
\begin{align}
\mathscr{K}_{\alpha_1\alpha_1',\alpha_2\alpha_2'} & = {\mathpzc G}_{\mu\nu}(k) [i\gamma_\mu]_{\alpha_1\alpha_1'} [i\gamma_\nu]_{\alpha_2\alpha_2'}\,,\\
 {\mathpzc G}_{\mu\nu}(k) & = \tilde{\mathpzc G}(k^2) T_{\mu\nu}(k)\,,
\end{align}
\end{subequations}
with $k^2 T_{\mu\nu}(k) = k^2 \delta_{\mu\nu} - k_\mu k_\nu$ and ($s=k^2$)
\begin{align}
\label{defcalG}
 \tfrac{1}{Z_2^2}\tilde{\mathpzc G}(s) & =
 \frac{8\pi^2 D}{\omega^4} e^{-s/\omega^2} + \frac{8\pi^2 \gamma_m \mathcal{F}(s)}{\ln\big[ \tau+(1+s/\Lambda_{\rm QCD}^2)^2 \big]}\,,
\end{align}
%KDinteractiondefcalG
where $\gamma_m=4/\beta_0$, $\beta_0=11 - (2/3)n_f$, $n_f=4$,
$\Lambda_{\rm QCD}=0.234\,$GeV,
$\tau={\rm e}^2-1$ $(\ln {\rm e} = 1)$,
and ${\cal F}(s) = \{1 - \exp(-s/[4 m_t^2])\}/s$, $m_t=0.5\,$GeV.
The development of Eqs.\,\eqref{KDinteraction}, \eqref{defcalG} is summarised in Ref.\,\cite{Qin:2011dd} and their connection with QCD is described in Ref.\,\cite{Binosi:2014aea}.  Some points may nevertheless bear repeating.  Namely, the interaction is deliberately consistent with that determined in studies of QCD's gauge sector and it preserves the one-loop renormalisation group behaviour of QCD.

$Z_2$ in Eq.\,\eqref{defcalG} is the dressed-quark wave function renormalisation constant.  We employ a mass-independent momentum-subtraction renormalisation scheme for the gap and inhomogeneous vertex equations, implemented by using the scalar WGT identity and fixing all renormalisation constants in the chiral limit \cite{Chang:2008ec}.  In the first applications of this DSE approach to hadron observables \cite{Frank:1995uk, Maris:1997tm} (and many that have followed), the renormalisation scale was chosen deep in the spacelike region: $\zeta=\zeta_{19} := 19\,$GeV, primarily to ensure simplicity in the nonperturbative renormalisation procedure.
This choice entails that the dressed quasiparticles obtained as solutions to the DSEs remain intact and thus serve as the dominant degrees-of-freedom for all observables.  This is adequate for infrared quantities, such as hadron masses: flexibility of model parameters and the bridge with QCD enable valid predictions to be made.  However, it generates errors in form factors and parton distributions.  With form factors, the correct power-law behaviour is obtained, but the scaling violations deriving from anomalous operator dimensions are wrong (see, \emph{e.g}.\ Ref.\cite{Maris:1998hc}); and for parton distributions, the natural connection between the renormalisation scale and the reference scale for evolution equations is lost, again because parton loops are suppressed when renormalising a RL truncation study at deep spacelike momenta so the computed anomalous dimensions are wrong.

As explained elsewhere \cite{Raya:2015gva, Gao:2017mmp, Ding:2018xwy}, the solution to these problems is to renormalise the DSE solutions at a typical hadronic scale, where the dressed quasiparticles are the correct degrees-of-freedom.  This recognises that a given meson's Poincar\'e covariant wave function and correlated vertices, too, must evolve with $\zeta$ \cite{Lepage:1979zb, Efremov:1979qk, Lepage:1980fj}.  Such evolution enables the dressed-quark and -antiquark degrees-of-freedom, in terms of which the wave function is expressed at a given scale $\zeta^2=Q^2$, to split into less well-dressed partons via the addition of gluons and sea quarks in the manner prescribed by QCD dynamics.  These effects are automatically incorporated in bound-state problems when the complete quark-antiquark scattering kernel is used; but aspects are lost when that kernel is truncated, and so it is with RL truncation.  We therefore renormalise our DSEs at the hadronic scale $\zeta=\zeta_H$.

A natural value for the hadronic scale, $\zeta_H$, must now be determined.  To that end, recall that QCD possesses a process-independent effective charge \cite{Binosi:2016nme, Rodriguez-Quintero:2018wma, Cui:2019dwv}: $\alpha_{\rm PI}(k^2)$.  This running-coupling saturates in the infrared: $\alpha_{\rm PI}(0)/\pi \approx 1$, owing to the dynamical generation of a gluon mass-scale \cite{Boucaud:2011ug, Aguilar:2015bud}.  These features and a smooth connection with pQCD (and hence Eq.\,\eqref{defcalG}) are expressed in the following algebraic expression:
\begin{align}
\label{alphaPI}
\alpha_{\rm PI}(k^2) = \frac{\pi \gamma_m  }{\ln[(m_\alpha^2+k^2)/\Lambda_{\rm QCD}^2]}\,,
\end{align}
$m_\alpha = 0.30\,$GeV$\,\gtrsim \Lambda_{\rm QCD}$.  Evidently, $m_\alpha$ is an essentially nonperturbative scale whose existence ensures that modes with $k^2 \lesssim m_\alpha^2$ are screened from interactions.  It therefore serves to define the natural boundary between soft and hard physics; hence, we identify
\begin{equation}
\label{setzetaH}
\zeta_H=m_\alpha\,.
\end{equation}

Returning to Eqs.\,\eqref{KDinteraction}, \eqref{defcalG}, computations \cite{Qin:2011dd, Qin:2011xq, Chen:2018rwz} reveal that observable properties of light-quark ground-state vector- and flavour-nonsinglet pseudoscalar-mesons are practically insensitive to variations of $\omega \in [0.4,0.6]\,$GeV, so long as
\begin{equation}
 \varsigma^3 := D\omega = {\rm constant}.
\label{Dwconstant}
\end{equation}
%(The midpoint $\omega=0.5\,$GeV is usually employed.)
This feature extends to numerous properties of the nucleon and $\Delta$-baryon \cite{Eichmann:2008ef, Eichmann:2012zz, Wang:2018kto, Qin:2019hgk}.  The value of $\varsigma$ is typically chosen to reproduce the measured value of the pion's leptonic decay constant, $f_\pi$.  In RL truncation, this requires
\begin{equation}
\label{varsigmalight}
\varsigma =0.82\,{\rm GeV,}
\end{equation}
with renormalisation-group-invariant current-quark mass
\begin{equation}
\hat m_u = \hat m_d = \hat m = 6.7\,{\rm MeV}\,,
\end{equation}
which corresponds to a one-loop evolved mass of $m^{\zeta_2} = 4.6\,$MeV.
%%Here is the parameters I used.
%  (a) Renormalisation-group-invariant current-quark mass: \hat m=7 MeV.
%  (b) One-loop evolved mass: m_{2GeV}=4.9 MeV.
%  (c) Decay constant: f_\pi=0.094 GeV.
%  (d) Pseudoscalar projections: \rho^{2GeV}_\pi=(0.425 GeV)^2.
In solving the DSEs relevant to pion physics, we will subsequently employ $\omega=0.5\,$GeV, the midpoint of the domain of insensitivity.

The next step on the way to obtaining ${\mathpzc q}^\pi(x;\zeta_H)$ is to perform a coupled solution of the dressed-quark gap-  and pion Bethe-Salpeter-equations, defined via Eqs.\,\eqref{KDinteraction}, \eqref{defcalG}, following Ref.\,\cite{Maris:1997tm} and adapting the algorithm improvements from Ref.\,\cite{Krassnigg:2009gd} when necessary.

\subsection{Mellin moments}
With $S$ and $\Gamma_\pi$ in hand, one can calculate the Mellin-moments:
\begin{subequations}
\label{MellinMoments}
\begin{align}
& \langle   x^m  \rangle_{\zeta_H}^\pi  = \int_0^1dx\, x^m {\mathpzc q}^\pi(x;\zeta_H) \label{MellinA}\\
& = \frac{N_c}{n\cdot P} {\rm tr}\! \int_{dk}\! \left[\frac{n\cdot k_\eta}{n\cdot P}\right]^m \Gamma_\pi(k_{\bar\eta},P)\, S(k_{\bar\eta})\, \nonumber \\
&\qquad  \qquad \qquad \times n\cdot\partial_{k_\eta} \left[ \Gamma_\pi(k_\eta,-P) S(k_\eta) \right];
\end{align}
\end{subequations}
and if enough of these moments can be computed, then they can be used to reconstruct the distribution.  Usefully, using Eq.\,\eqref{xoneminusx}, one finds that the value of any given odd moment, $\langle x^{m_{\rm o}}\rangle_{\zeta_H}^{\pi}$, $m_{\rm o}=2 \bar m +1$, $\bar m \in \mathbb Z$, is known once all lower even moments are computed, \emph{e.g}.:
\begin{subequations}
\begin{align}
\langle x  \rangle_{\zeta_H}^\pi & = \frac{1}{2} \langle x^0  \rangle_{\zeta_H}^\pi = \frac{1}{2} \,,\\
\langle x^3  \rangle_{\zeta_H}^\pi & =-\frac{1}{4} \langle x^0  \rangle_{\zeta_H}^\pi+ \frac{3}{2} \langle x^2  \rangle_{\zeta_H}^\pi\,,\label{mom3}\\
% x0/2 - (5*x2)/2 + (5*x4)/2
\langle x^5  \rangle_{\zeta_H}^\pi & =\frac{1}{2} \langle x^0  \rangle_{\zeta_H}^\pi - \frac{5}{2} \langle x^2  \rangle_{\zeta_H}^\pi + \frac{5}{2} \langle x^4  \rangle_{\zeta_H}^\pi \,, \label{mom5}\\
\langle x^7  \rangle_{\zeta_H}^\pi & = -\frac{17}{8} \langle x^0  \rangle_{\zeta_H}^\pi
+ \frac{21}{2}\langle x^2  \rangle_{\zeta_H}^\pi \nonumber \\
& \qquad - \frac{35}{4}\langle x^4  \rangle_{\zeta_H}^\pi
+ \frac{7}{2}\langle x^6  \rangle_{\zeta_H}^\pi\,. \label{mom7}
%
%(-17*x0)/8 + (21*x2)/2 - (35*x4)/4 + (7*x6)/2
\end{align}
\end{subequations}
Such identities can be used to validate any numerical method for computing the moments defined by Eq.\,\eqref{MellinMoments}.

Every moment defined by Eq.\,\eqref{MellinMoments} is finite.  However, direct calculation of these moments using numerically determined inputs for the propagator and Bethe-Salpeter amplitude is difficult in practice owing to an amplification of oscillations produced by the $[n\cdot k_\eta]^m$ factor: in any perfect procedure, the oscillations cancel; but that is hard to achieve numerically.  We therefore introduce a convergence-factor\footnote{Owing to the nature of the integrand, the convergence factor can be omitted for $m\leq 2$: it only plays a role for $m\geq 3$.}
\begin{equation}
\label{Cfactor}
{\mathpzc C}_m(k^2 r^2)= 1/[1+k^2 r^2]^{m/2}:
\end{equation}
the moment is computed as a function of $r^2$; and the final value is obtained by extrapolation to $r^2=0$.
This procedure is reliable for the lowest six moments, $m=0,1,\ldots,5$ \cite{Li:2016dzv}.  According to Eq.\,\eqref{mom5}, the $m=5$ moment is not independent; but its direct calculation enables one to ensure that the lower even moments are correct.

One can extend this set of moments by using the Schlessinger point method (SPM) \cite{Schlessinger:1966zz, PhysRev.167.1411, Tripolt:2016cya, Chen:2018nsg, Binosi:2019ecz} to construct an analytic function, $M_S(z)$, whose values at $z=0,1,\ldots,5$ agree with the moments computed directly and for which $M_S(7)$ satisfies Eq.\,\eqref{mom7} when $M_S(0)$, $M_S(2)$, $M_S(4)$, $M_S(6)$ are used for the even moments.  The function $M_S(z)$ then provides an estimate for all moments of the distribution, which is exact for $m\leq 5$.

%It is here worth observing that the SPM is based on the Pad\'e approximant.  It is able to accurately reconstruct a function in the complex plane within a radius of convergence specified by that one of the function's branch points which lies nearest to the real domain from which the sample points are drawn.  Moreover, owing to the procedure's discrete nature, the reconstruction may also provide a reasonable continuation on a larger domain, but this cannot be guaranteed and, hence, each case must be treated individually.

%%our S and Gamma by Eq.(1), rho and parameters chosen from Eq.(14) and (17).

\begin{figure}[t]
\centerline{%
\includegraphics[clip, width=0.42\textwidth]{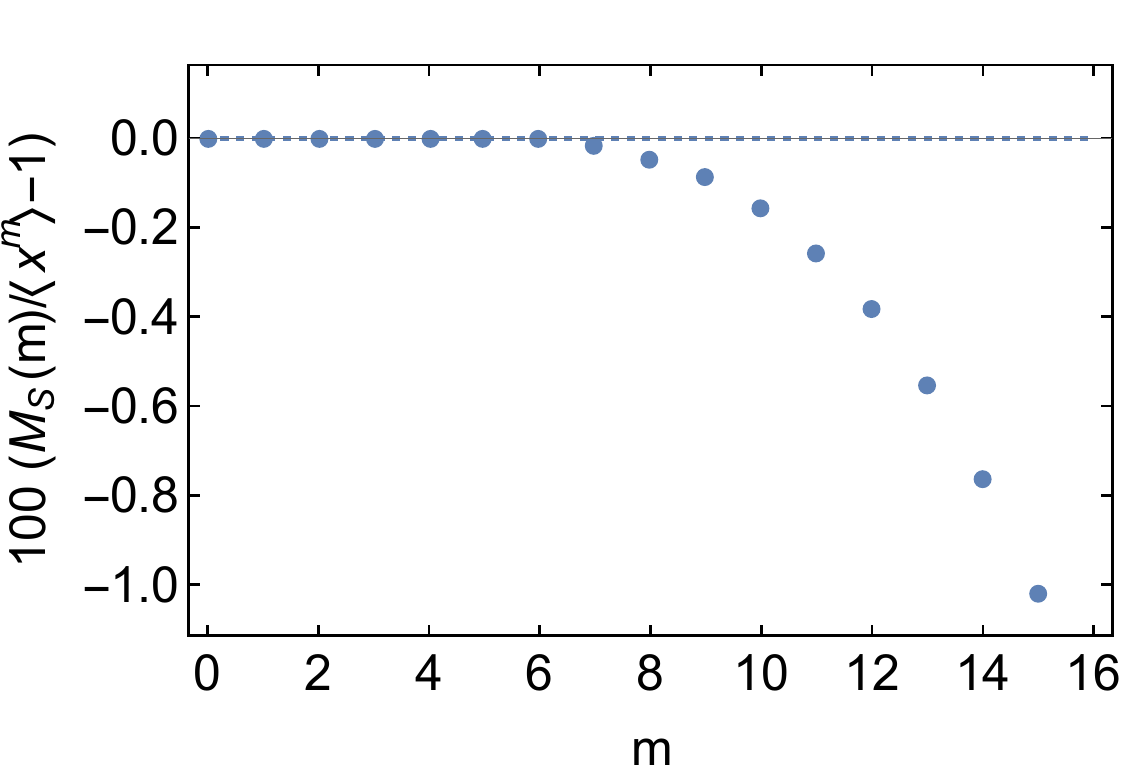}}
\caption{\label{SPMcompare}
Comparison between moments evaluated using the SPM approximation in Eq.\,\eqref{SPMtest} with those computed directly using the algebraic model in Ref.\,\cite{Xu:2018eii}: the magnitude of the relative error is $<0.1\,$\% for $m\leq 10$.}
\end{figure}

We illustrate the efficacy of the SPM approach using the nontrivial algebraic model described in Ref.\,\cite{Xu:2018eii} (Eqs.\,(1), (14), (17) and Sec.IV.A).  We computed fifty Mellin moments directly, then used  the first six moments and the procedure described above to obtain the following SPM approximation:
\begin{align}
\label{SPMtest}
M_S(z) = \frac{a_0 + a_1 z + a_2 z^2}{a_0 + b_1 z + b_2 z^2 + b_3 z^3}\,,
\end{align}
with the coefficients specified in Table~\ref{tabSPMcoefficients}.  Figure~\ref{SPMcompare} compares the moments obtained using the SPM approximation with the true moments: the magnitude of the relative error is $<0.2\,$\% for $m\leq 10$ and $<1$\% for $m\leq 15$, \emph{i.e}.\ the SPM produces accurate approximations to the first sixteen moments, working with just six.  The relative error for the fiftieth moment is $-48$\%.

\begin{table}[t]
\caption{\label{tabSPMcoefficients}
Computed coefficients for the SPM approximation to the Mellin moments of the pion valence-quark distribution function, Eq.\,\eqref{SPMtest}.
\emph{Upper panel} -- algebraic inputs for the propagator and Bethe-Salpeter amplitude.
\emph{Lower panel} -- realistic inputs: column 1, using $m=0,\ldots,3$; and column~2, $m=0,\ldots,5$.
}
\begin{center}
\begin{tabular}%{|c|c|c|c|c|c|c|}\hline
%{\hsize}
{
c@{\extracolsep{0ptplus1fil}}|
c@{\extracolsep{0ptplus1fil}}}\hline\hline
          & algebraic  \\\hline
$a_0\ $ &$\phantom{-}22.15848512824146\phantom{0}$ \\
$a_1\ $ &$\phantom{-2}6.882746278686694$  \\
$a_2\ $ &$ \phantom{8}-0.1087281502480409$  \\
$b_1\ $ & $\phantom{-}25.18581743280522\phantom{0}$ \\
$b_2\ $ & $\phantom{-2}9.520703952313553 $ \\
$b_3\ $ & $\phantom{-2}1\phantom{.520703952313553}$ \\\hline\hline
\end{tabular}
\end{center}
\smallskip

\begin{center}
\begin{tabular*}%{|c|c|c|c|c|c|c|}\hline
{\hsize}
{
l@{\extracolsep{0ptplus1fil}}|
c@{\extracolsep{0ptplus1fil}}
c@{\extracolsep{0ptplus1fil}}}\hline\hline
          & realistic: $0-3$  & realistic: $0-5$ \\\hline
$a_0\ $ &$\phantom{-}6\phantom{.05050505050504085}$
    & $\phantom{-}24.45939048190962\phantom{0}$ \\
$a_1\ $ &$\phantom{-}0.05050505050504085$
    & $\phantom{-}16.78622673154534\phantom{0}$ \\
$a_2\ $ &$ \phantom{-}0\phantom{0.8585917712408488}$
    & $\phantom{8}- 0.8585917712408488$ \\
$b_1\ $ & $\phantom{-}5.101010101010094\phantom{85}$
    & $\phantom{-}36.46584243665940\phantom{0}$ \\
$b_2\ $ & $\phantom{-}1\phantom{.05050505050504085} $
    & $\phantom{-}18.84881796585922 \phantom{0}$ \\
$b_3\ $
    & $\phantom{-}0\phantom{.05050505050504085}$
    & $\phantom{-18}1\phantom{.0505050505050408}$\\
\hline\hline
\end{tabular*}
\end{center}
\end{table}
% -true algebraic model
%(22.15848512824146 + 6.882746278686694 z - 0.1087281502480409 z^2)/(22.15848512824146 + 25.18581743280522 z + 9.520703952313553 z^2 + 1.000000000000000 z^3)
 %%
%(26.94891453474296 + 8.038971576939086 z - 0.1051988919337212 z^2)/(26.94891453474296 + 30.56753912839169 z + 11.24892077636199 z^2 + 1.000000000000000 z^3)
%
%(26.482758620586424` + 8.338137931030182` z - 0.13124137931004487` z^2)/(26.482758620586424` + 30.448275861978047` z + 11.44827586204865` z^2 + 1.` z^3)

Having validated the SPM, we computed the moments in Eq.\,\eqref{MellinMoments} for $m=0,1,\ldots, 5$ using the numerical results for $S$ and $\Gamma_\pi$ obtained with the DSE kernels specified by Eqs.\,\eqref{KDinteraction}, \eqref{defcalG}.\footnote{We used the SPM to assist with extrapolation $r^2 \to 0$, Eq.\,\eqref{Cfactor}, and Eqs.\,\eqref{mom3}, \eqref{mom5} to check our results.}
Then, to compensate for potential propagation of numerical quadrature error in the moment computations, we constructed two SPM approximations to the results: one based on the $m=0,1,2,3$ four-element subset; and another using the complete set of six moments.  In each case, the collection of moments is described by a function of the form in Eq.\,\eqref{SPMtest} with the coefficients in the lower panel of Table~\ref{tabSPMcoefficients}.

Working with the first eleven SPM-approximant moments in each case, we reconstructed a pion valence-quark distribution; and subsequently defined our result to be the average of these functions:
\begin{align}
\nonumber {\mathpzc q}^\pi&(x;\zeta_H)  = 213.32 \, x^2 (1-x)^2\\
& \quad \times  [1 - 2.9342  \sqrt{x(1-x)} + 2.2911 \,x (1-x)]\,. \label{qpizetaH}
\end{align}
%253.62003173893243*(1 - x)^2*x^2*(1 + 2.6974201897292325*(1 - x)*x - 3.171989803894321*Sqrt[(1 - x)*x])
%213.31662630896608*(1 - x)^2*x^2*(1 + 2.291114178403205*(1 - x)*x - 2.934232803738537*Sqrt[(1 - x)*x])
%
Notably, the $x\simeq 0,1$ endpoint behaviour of the hadron-scale distribution is completely fixed by algebraic analysis \cite{Chang:2014lva}: as shown, the exponent is ``2'', reproducing Eq.\,\eqref{PDFQCD}.  This feature removes the need to use moments of arbitrarily high order, enabling one to focus instead on the lower-order moments which provide information on the mid-$x$ shape.

\begin{figure}[t]
\centerline{%
\includegraphics[clip, width=0.42\textwidth]{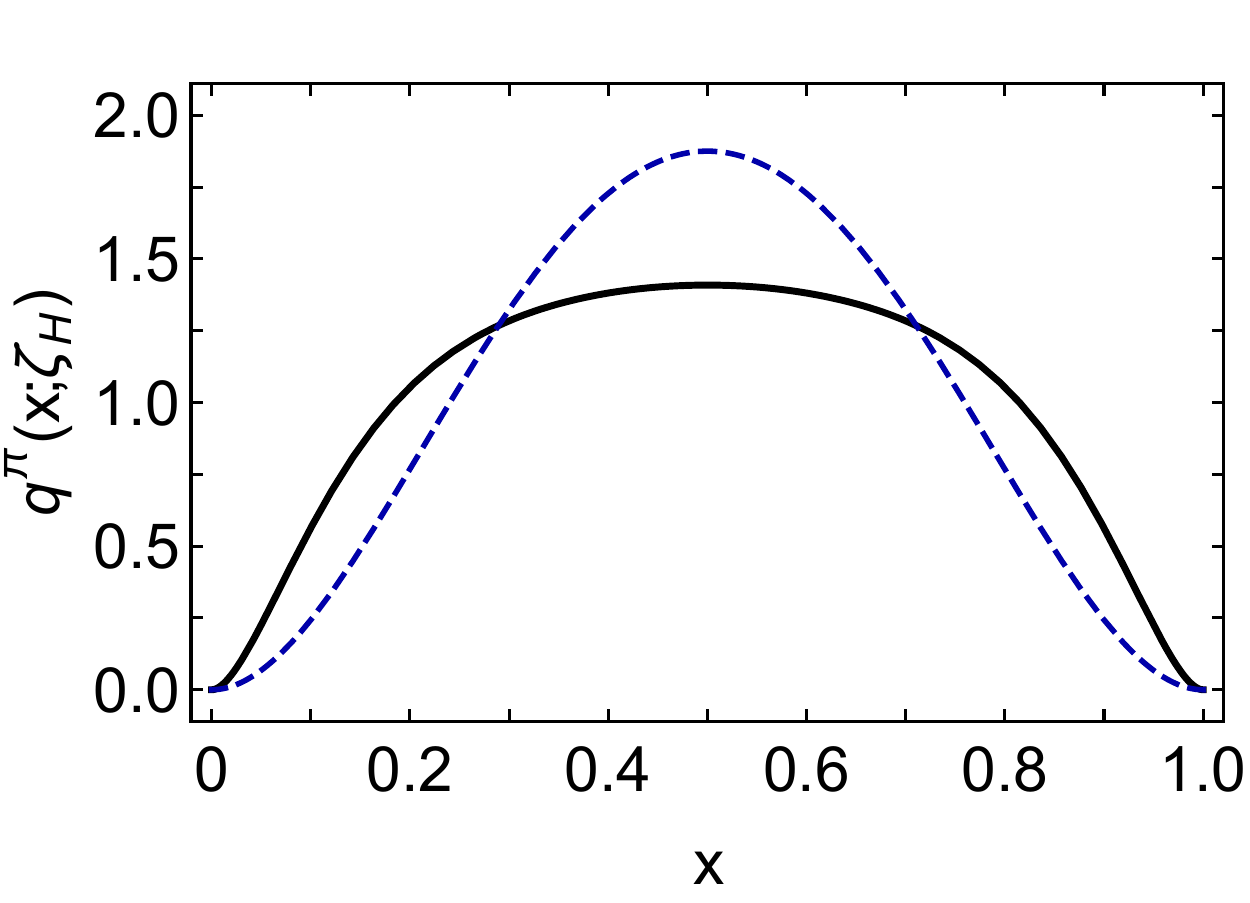}}
\caption{\label{plotqpizetaH}
Solid (black) curve: pion valence-quark distribution function at the hadronic scale, $\zeta_H$, Eq.\,\eqref{qpizetaH}.
Dashed (blue) curve: scale-free distribution, Eq.\,\eqref{qpiscalefree}.
}
\end{figure}

One remark may be valuable here.  This application of the SPM requires the coefficient of the highest active denominator power in Eq.\,\eqref{SPMtest} to be unity.  Hence, when one uses Eq.\,\eqref{SPMtest} for $m=0,1,2,3$ moments, $b_2=1$ and $a_2=0=b_3$.  Referring to the lower panels of Table~\ref{tabSPMcoefficients}, this presents an appearance of sensitivity in the coefficients to the number of moments employed; but that is misleading.  The relevant measure is not these coefficients, but the similarity between the curves obtained via reconstruction.
Our result, Eq.\,\eqref{qpizetaH}, is depicted in Fig.\,\ref{plotqpizetaH}.  The mean absolute relative error between its first eleven moments and those of the separate reconstructed distributions is 4(3)\%.
%
%%reproduces the computed values of the first ten nontrivial moments with mean absolute relative difference $0.014(9)$\%, \emph{i.e}.\ at the level of one-part-in-ten-thousand.

Given the remarks in Sec.\,\ref{Introduction}, it is worth reiterating that Eq.\,\eqref{qpizetaH} exhibits the $x\simeq 1$ behaviour predicted by the QCD parton model, Eq.\,\eqref{PDFQCD}; and because it is a purely valence distribution, this same behaviour is also evident on $x\simeq 0$.
However, in contrast to the scale-free valence-quark distribution computed in Ref.\,\cite{Chang:2014lva}:
\begin{equation}
\label{qpiscalefree}
q_{\rm sf}(x) \approx 30 \, x^2 (1-x)^2,
\end{equation}
obtained using parton-model-like algebraic representations of $S$, $\Gamma_\pi$, the distribution computed with realistic inputs is a much broader function.  A similar effect is observed in the pion's leading-twist valence-quark distribution amplitude \cite{Chang:2013pq} and those of other mesons \cite{Gao:2014bca, Shi:2015esa, Gao:2016jka, Li:2016dzv, Li:2016mah}.  The cause is the same, \emph{viz}.\ the valence-quark distribution function is hardened owing to DCSB, which is a realisation of the mechanism responsible for the emergence of mass in the Standard Model \cite{Roberts:2016vyn}.  Emergent mass is expressed in the momentum-dependence of all QCD Schwinger functions.  It is therefore manifest in the pointwise behaviour of wave functions, elastic and transition form factors, \emph{etc}.; and as we have now displayed, also in parton distributions.  (This was to be expected, given the connection between light-front wave functions and parton distributions.)

\section{Evolution of pion distribution functions}
\label{secEvolution}
The pion valence-quark distribution in Eq.\,\eqref{qpizetaH} is computed at $\zeta_H=m_\alpha$, Eq.\,\eqref{setzetaH}.  On the other hand, existing lQCD calculations of low-order moments \cite{Best:1997qp, Detmold:2003tm, Brommel:2006zz, Oehm:2018jvm} and phenomenological fits to pion parton distributions are typically quoted at $\zeta \approx \zeta_2=2\,$GeV \cite{Sutton:1991ay, Gluck:1999xe, Barry:2018ort}; and the scale relevant to the E615 data is $\zeta_5=5.2\,$GeV \cite{Conway:1989fs, Wijesooriya:2005ir}.
We therefore employ the effective charge in Eq.\,\eqref{alphaPI} to integrate the one-loop DGLAP equations, therewith evolving ${\mathpzc q}^\pi(x;\zeta_H=m_\alpha)$ to obtain results for ${\mathpzc q}^\pi(x;\zeta_2)$ and ${\mathpzc q}^\pi(x,\zeta_5)$.
This procedure ensures that saturation of the effective charge is expressed, \emph{e.g}.\ $\alpha_{\rm PI}(\zeta_H)/(2\pi)=0.20$, $[\alpha_{\rm PI}(\zeta_H)/(2\pi)]^2=0.04$, stabilising our evolved results on $\zeta > \zeta_H$.
Notably, given that $\zeta_H=m_\alpha$ is fixed by our analysis, all results are predictions.
We checked that with fixed $\zeta_H$, varying $m_\alpha \to (1 \pm 0.1) m_\alpha$ does not measurably affect the evolved distributions.  We therefore report results with $m_\alpha$ fixed and an uncertainty determined by varying $\zeta_H \to (1\pm0.1) \zeta_H$.

\begin{table}[b]
\caption{\label{fittingparameters}
Coefficients and powers that reproduce the computed pion valence-quark distribution functions, depicted in Figs.~\ref{qpizeta2}, \ref{qpizeta5}, when used in Eq.\,\eqref{PDFform}.
}
\begin{center}
\begin{tabular*}%{|c|c|c|c|c|c|c|}\hline
{\hsize}
{
l@{\extracolsep{0ptplus1fil}}|
c@{\extracolsep{0ptplus1fil}}
c@{\extracolsep{0ptplus1fil}}
c@{\extracolsep{0ptplus1fil}}
c@{\extracolsep{0ptplus1fil}}
c@{\extracolsep{0ptplus1fil}}}\hline\hline
 & ${\mathpzc n}_{{\mathpzc q}^\pi}\ $ & $\alpha\ $ & $\beta\ $ & $\rho\ $ & $\gamma\ $ \\\hline
             % & $10.98\ $ & $-0.052\ $ & $2.29\ $ & $-1.40\ $ & $0.637\ $  \\
             & $9.83\ $ & $-0.080\ $ & $2.29\ $ & $-1.27\ $ & $0.511\ $  \\
%$\zeta_2\ $ & $\phantom{1}9.26\ $ & $-0.096\ $ & $2.37\ $ & $-1.32\ $ & $0.594\ $ \\
$\zeta_2\ $ & $8.31\ $ & $-0.127\ $ & $2.37\ $ & $-1.19\ $ & $0.469\ $ \\
            % & $\phantom{1}7.38\ $ & $-0.140\ $ & $2.44\ $ & $-1.21\ $ & $0.538\ $ \\\hline
             & $7.01\ $ & $-0.162\ $ & $2.47\ $ & $-1.12\ $ & $0.453\ $ \\\hline
             & $7.81\ $ & $-0.153\ $ & $2.54\ $ & $-1.20\ $ & $0.505\ $ \\
$\zeta_5\ $ & $7.28\ $ & $-0.169\ $ & $2.66\ $ & $-1.21\ $ & $0.531\ $ \\
             & $6.48\ $ & $-0.188\ $ & $2.78\ $ & $-1.19\ $ & $0.555\ $  \\
\hline\hline
\end{tabular*}
\end{center}
\end{table}
%fpdfFM\[Zeta]5store = (7.698259159294685*(1 + (0.6234795897702227*(1 - x)^1.3205213906392288)/        x^0.07519328462674237 - (1.3032532992534949*(1 - x)^0.6602606953196144)/x^0.03759664231337118)*(1 -        x)^2.6410427812784576)/x^0.15038656925348473;
%{\[Alpha] -> -0.150387, \[Rho] -> -1.30325, \[Gamma] -> 0.62348}
%
% (10.98246491543057*(1 + (0.6373037719153538*(1 - x)^1.146070354541323)/x^0.026092415839046233 - (1.4029515105830297*(1 - x)^0.5730351772706616)/x^0.013046207919523116)* (1 - x)^2.292140709082646)/x^0.052184831678092465
%
% (9.26197822786096*(1 + (0.5935607407148072*(1 - x)^1.183490105663935)/x^0.04790244906426601 - (1.324668702508156*(1 - x)^0.5917450528319675)/x^0.023951224532133004)*  (1 - x)^2.36698021132787)/x^0.09580489812853202
%(7.38264935284909*(1 + (0.5384359427489233*(1 - x)^1.2218641911132788)/x^0.07022994791683992 - (1.214442067834761*(1 - x)^0.6109320955566394)/x^0.03511497395841996)*  (1 - x)^2.4437283822265576)/x^0.14045989583367985  \[Rho] -> -1.21444, \[Gamma] -> 0.538436

\subsection{$\zeta_H \to \zeta_2$}
\begin{figure}[t]
\centerline{%
\includegraphics[clip, width=0.42\textwidth]{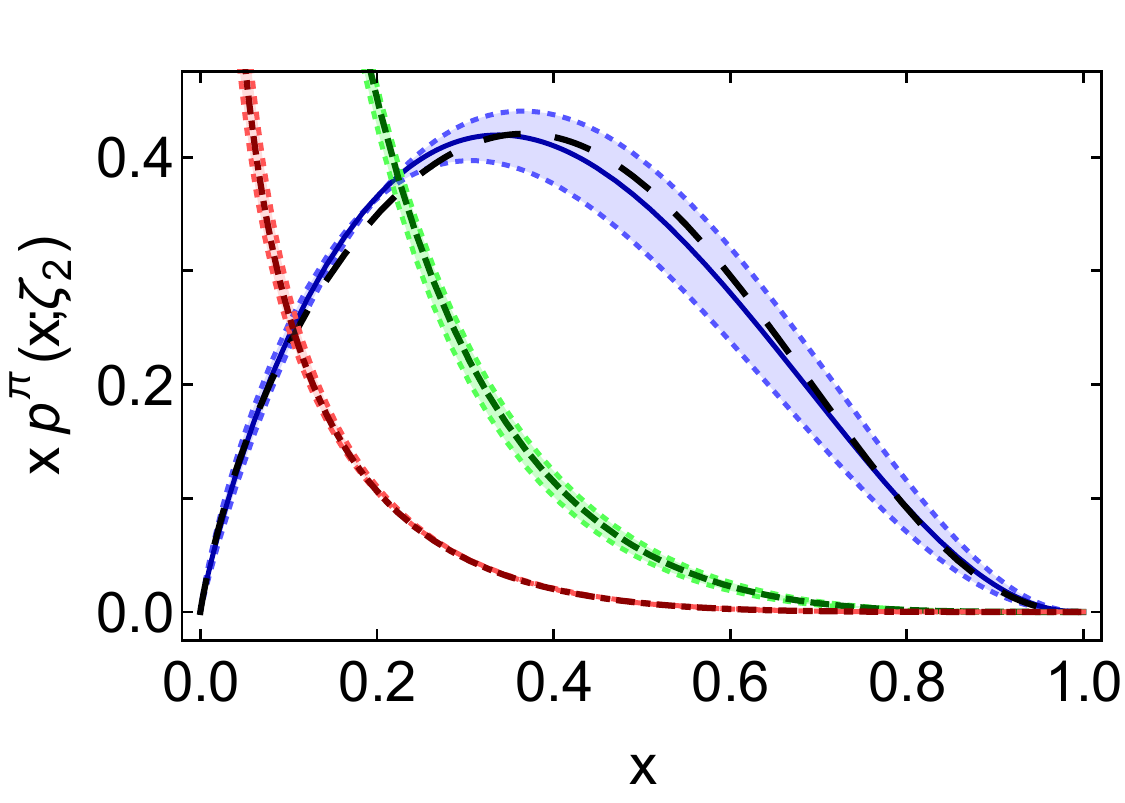}}
\caption{\label{qpizeta2}  Pion valence-quark momentum distribution function, $x p^\pi(x;\zeta)$, $p={\mathpzc q}$, evolved $\zeta_H \to \zeta_2=2\,$GeV -- solid (blue) curve embedded in shaded band; and long-dashed (black) curve -- $\zeta_2$ result from Ref.\,\cite{Hecht:2000xa}.
Eqs.\,\eqref{seaglueFunction}, \eqref{seagluefit}:
gluon momentum distribution in pion, $x g^\pi(x;\zeta_2)$ -- dashed (green) curve within shaded band;
and sea-quark momentum distribution, $x S^\pi(x;\zeta_2)$ -- dot-dashed (red) curve within shaded band.
In all cases, the shaded band indicates the effect of $\zeta_H \to \zeta_H (1 \pm 0.1)$.
}
\end{figure}

Our prediction for ${\mathpzc q}^\pi(x;\zeta_2)$ is depicted in Fig.\,\ref{qpizeta2}.  The solid curve and surrounding bands are described by the following function, a generalisation of Eq.\,\eqref{qpizetaH}:
\begin{align}
\nonumber {\mathpzc q}^\pi(x) & = {\mathpzc n}_{{\mathpzc q}^\pi} \,x^\alpha (1-x)^\beta \\
& \times [1 + \rho\, x^{\alpha/4} (1-x)^{\beta/4} + \gamma \,x^{\alpha/2} (1-x)^{\beta/2} ]\,,
\label{PDFform}
\end{align}
%{\[Alpha] -> -0.10164, \[Beta] ->  2.35649, \[Rho] -> -1.30584, \[Gamma] -> 0.584702}
where ${\mathpzc n}_{{\mathpzc q}^\pi}$ ensures Eq.\,\eqref{BaryonSum} and the powers and coefficients are listed in Table~\ref{fittingparameters}.  Evidently, the large-$x$ exponent is
\begin{equation}
\beta(\zeta_2) = 2.38(9)\,.
\end{equation}

Here it is also worth listing an array of associated, calculated low-order moments in comparison with those obtained in the more recent lQCD simulations:
\begin{equation}
\label{momentslQCD}
\begin{array}{l|lll}
\zeta_2  & \langle x \rangle_u^\pi & \langle x^2 \rangle_u^\pi & \langle x^3 \rangle_u^\pi\\\hline
%\mbox{\cite{Best:1997qp}} & 0.28(8) & 0.11(3) & 0.048(20)\\
\mbox{Ref.\,\cite{Detmold:2003tm}} & 0.24(2) & 0.09(3) & 0.053(15)\\
\mbox{Ref.\,\cite{Brommel:2006zz}} & 0.27(1) & 0.13(1) & 0.074(10)\\
\mbox{Ref.\,\cite{Oehm:2018jvm}} & 0.21(1) & 0.16(3) & \\
\mbox{Ref.\,\cite{Joo:2019bzr}} & 0.254(03) & 0.094(12) & 0.057(04) \\\hline
%{\rm average} & 0.26(8) & 0.11(4) & 0.058(27)\\\hline
%{\rm average} & 0.24(2) & 0.13(4) & 0.064(18)\\\hline
{\rm average} & 0.24(2) & 0.119(18) & 0.061(06)\\\hline
{\rm Herein} & 0.24(2) & 0.098(10) & 0.049(07)
% {\rm herein} & 0.26 & 0.11 & 0.052 ... Chen
\end{array}\,.
\end{equation}
%<x>= 0.2075(53)stat(20)sys(90)Z a
% <x2> = 0.163(23)stat(25)sys
%
%{1, 0.259544}, {2, 0.108411}, {3, 0.0560606},
%{1, 0.224122}, {2, 0.0861982}, {3, 0.0420332},
Both continuum and lQCD results agree on the light-front momentum fraction carried by valence-quarks in the pion at $\zeta=\zeta_2$:
\begin{equation}
\langle 2 x \rangle_{\mathpzc q}^\pi =0.48(3)\,,
\end{equation}
\emph{i.e}.\ roughly one-half.  This is consistent with a recent phenomenological analysis of data on $\pi$-nucleus Drell-Yan and leading neutron electroproduction \cite{Barry:2018ort}:  $\langle 2 x \rangle_{\mathpzc q}^\pi  = 0.48(1)$ at $\zeta=2.24\,$GeV.

As explained above, the pion is purely a bound-state of a dressed-quark and dressed-antiquark at the hadronic scale, $\zeta_H$.  Sea and glue distributions are zero at $\zeta_H$.  They are generated by QCD evolution on $\zeta>\zeta_H$.  Using LO evolution with the coupling in Eq.\,\eqref{alphaPI} we obtain the sea and glue distributions in Fig.\,\ref{qpizeta2}, from which one computes the following momentum fractions ($\zeta=\zeta_2$):
\begin{equation}
\langle x\rangle^\pi_g = 0.41(2)\,, \quad
\langle x\rangle^\pi_{\rm sea} = 0.11(2)\,.
\end{equation}
The ordering of these values agrees with that in \cite{Barry:2018ort}, but our gluon momentum-fraction is $\sim 20$\% larger and that of the sea is commensurately smaller.

Our computed glue and sea momentum distributions are fairly approximated using the following simple functional form:
\begin{equation}
\label{seaglueFunction}
x p^\pi(x;\zeta) = {\mathpzc A} \, x^\alpha \, (1-x)^\beta\,,
\end{equation}
with the coefficient and powers listed here ($p=g=\,$glue, $p=S=\,$sea):
\begin{equation}
\label{seagluefit}
\begin{array}{l|cccc}
             & p & {\mathpzc A} & \alpha & \beta \\\hline
\zeta_2 &  g & 0.40 \mp 0.03 & -0.55 \mp 0.03 & 3.47 \pm 0.13 \\
          &  S & 0.13 \mp 0.01 & -0.53 \mp 0.05 & 4.51 \pm 0.03 \\\hline
\zeta_5 &  g & 0.34 \mp 0.04 & -0.62 \mp 0.04 & 3.75 \pm 0.12 \\
          &  S & 0.12 \pm 0.02 & -0.61 \mp 0.07 & 4.77 \pm 0.03 \\\hline
\end{array}\,.
\end{equation}

%\begin{subequations}
%\begin{align}
%x g^\pi(x;\zeta_2) & = 0.40_{+0.14}^{-0.14} \, (1-x)^{3.45}/x^{0.55}\,,\\
%x S^\pi(x;\zeta_2) & = 0.19_{+0.045}^{-0.045} \, (1-x)^{4.91}/x^{0.42}\,.
%\end{align}
%\end{subequations}

\subsection{$\zeta_H \to \zeta_5$}
\begin{figure}[t]
\centerline{%
\includegraphics[clip, width=0.42\textwidth]{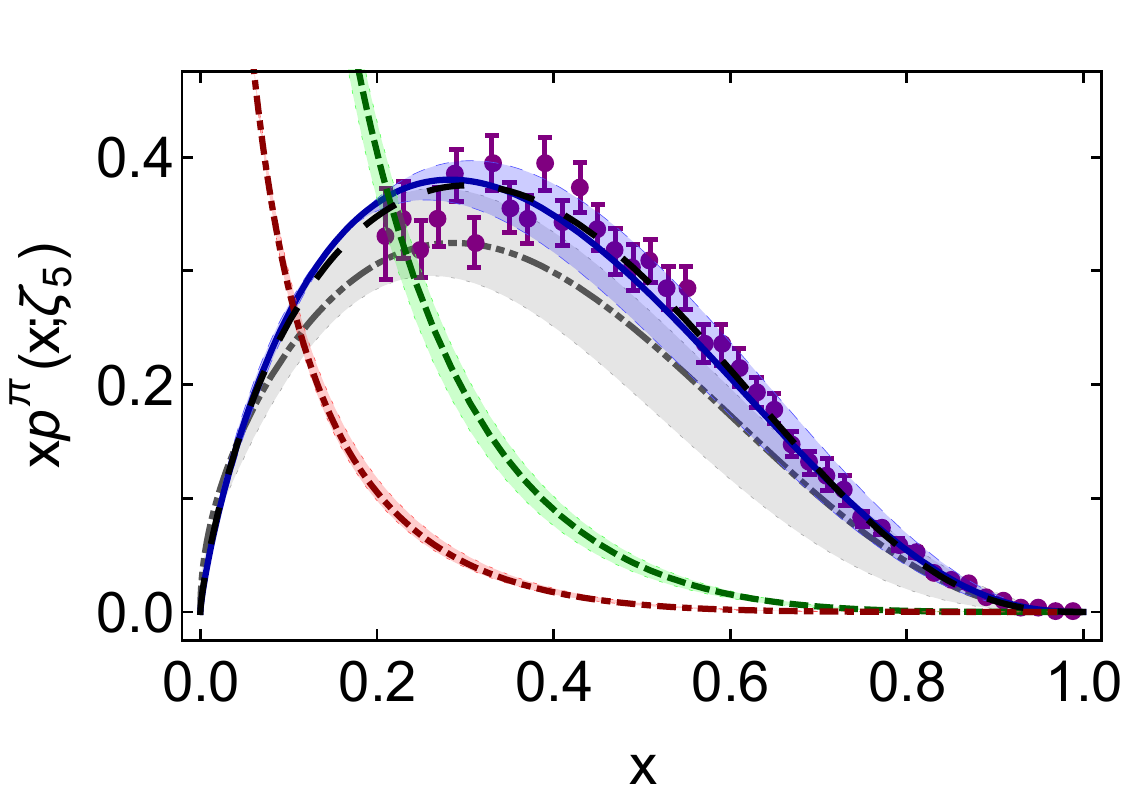}}
\caption{\label{qpizeta5}
Pion valence-quark momentum distribution function, $x {\mathpzc q}^\pi(x;\zeta)$, evolved $\zeta_H \to \zeta_5=5.2\,$GeV -- solid (blue) curve embedded in shaded band; and long-dashed (black) curve -- $\zeta_5$ result from Ref.\,\cite{Hecht:2000xa}.
Gluon momentum distribution in pion, $x g^\pi(x;\zeta_2)$ -- dashed (green) curve within shaded band;
and sea-quark momentum distribution, $x S^\pi(x;\zeta_2)$ -- dot-dashed (red) curve within shaded band.
See Eqs.\,\eqref{seaglueFunction}, \eqref{seagluefit}.
In all the above cases, the shaded band indicates the effect of $\zeta_H \to \zeta_H (1 \pm 0.1)$.
Dot-dot-dashed (grey) curve within shaded band -- lQCD result \cite{Sufian:2019bol}.
Data (purple) from Ref.\,\cite{Conway:1989fs}, rescaled according to the analysis in Ref.\,\cite{Aicher:2010cb}.
}
\end{figure}

Our predictions for the pion parton distributions at a scale relevant to the E615 experiment, \emph{i.e}.\ $\zeta_5=5.2\,$GeV \cite{Conway:1989fs, Wijesooriya:2005ir}, are depicted in Fig.\,\ref{qpizeta5}.  The solid curve and surrounding bands are described by the function in Eq.\,\eqref{PDFform} with the powers and coefficients listed in Table~\ref{fittingparameters}.  Evidently, the large-$x$ exponent is
\begin{equation}
\beta(\zeta_5) = 2.66(12)\,.
\end{equation}
Working with results obtained in an exploratory lQCD calculation \cite{Sufian:2019bol}, one finds $\beta_{\rm lQCD}(\zeta_5) = 2.45(58)$; and also the following comparison between low-order moments:
\begin{equation}
\label{momentslQCD5}
\begin{array}{l|lll}
\zeta_5  & \langle x \rangle_u^\pi & \langle x^2 \rangle_u^\pi & \langle x^3 \rangle_u^\pi\\\hline
%\mbox{\cite{Best:1997qp}} & 0.28(8) & 0.11(3) & 0.048(20)\\
\mbox{Ref.\,\cite{Sufian:2019bol}} & 0.17(1) & 0.060(9) & 0.028(7)\\
{\rm Herein} & 0.21(2) & 0.076(9) & 0.036(5)
% {\rm herein} & 0.26 & 0.11 & 0.052 ... Chen
\end{array} \,.
\end{equation}

The data in Fig.\,\ref{qpizeta5} is that reported in Ref.\,\cite{Conway:1989fs}, rescaled according to the analysis in Ref.\,\cite{Aicher:2010cb}.  Our prediction agrees with the rescaled data.  Importantly, no parameters were varied in order to achieve this outcome.

As above, the predictions for the glue and sea distributions in Fig.\,\ref{qpizeta5} were obtained using  LO evolution from $\zeta_H=m_\alpha \to \zeta_5$ with the coupling in Eq.\,\eqref{alphaPI}; and from these distributions one obtains the following momentum fractions ($\zeta=\zeta_5$):
\begin{equation}
\langle x\rangle^\pi_g = 0.45(1)\,, \quad
\langle x\rangle^\pi_{\rm sea} = 0.14(2)\,.
\end{equation}
The glue and sea momentum distributions are fairly described by the function in Eq.\,\eqref{seaglueFunction} evaluated using the coefficient and powers in the lower rows of Eq.\,\eqref{seagluefit}.\footnote{Recall that in the neighbourhood $\Lambda_{\rm QCD}^2/\zeta^2 \simeq 0$, for any hadron \cite{Altarelli:1981ax}: $\langle x\rangle_q =0$, $\langle x\rangle_g = 4/7\approx 0.57$, $\langle x\rangle_S = 3/7\approx 0.43$.}
%% https://arxiv.org/pdf/1608.02771.pdf  Eqs. (42), (43)
%% <x>q = 3 nf /(2 ng + 3 nf)
%% <x>g = 2 ng /(2 ng + 3 nf)
%% Altarelli:1981ax (4.82)-(4.85)
%% C2G = 3
%% TR = No.Flav./2 = 2 for nf=4
%% C2R = 4/3
%% sea = TR/(2. C2R + TR)
%% glue = 2 C2R/(2. C2R + TR)

Figure~\ref{qpizeta5} also displays the lQCD result for the pion valence-quark distribution function \cite{Sufian:2019bol} evolved to the E615 scale: dot-dot-dashed (grey) curve within bands.  As could be  anticipated from the comparisons listed in connection with Eq.\,\eqref{momentslQCD5}, the pointwise form of the lQCD prediction agrees with our result (within errors).  This is significant: two disparate treatments of the pion bound-state problem have now arrived at the same prediction for the pion's valence-quark distribution function.

\section{Summary and Perspective}
\label{epilogue}
Using a continuum approach to the two valence-body bound-state problem in quantum field theory, we presented a symmetry-preserving calculation of the pion's valence-quark distribution function, ${\mathpzc q}^\pi(x;\zeta_H)$, $\zeta_H$ is the hadronic scale [Sec.\,\ref{secSymmetries}]; and thereby unified the result with kindred predictions for the electromagnetic pion elastic and transition form factors
\cite{Chang:2013nia, Raya:2015gva, Raya:2016yuj, Gao:2017mmp, Chen:2018rwz, Ding:2018xwy} and numerous other observables (\emph{e.g}.\ Refs.\,\cite{Binosi:2018rht, Qin:2019hgk}).
Within this framework, the pion is purely a bound-state of a dressed-quark and dressed-antiquark at $\zeta_H$; consequently,
\begin{equation}
{\mathpzc q}^\pi(x;\zeta_H)={\mathpzc q}^\pi(1-x;\zeta_H)\,.
\end{equation}
Capitalising on this, we directly computed the first three independent Mellin moments of ${\mathpzc q}^\pi(x;\zeta_H)$ and therefrom
%, using the Schlessinger point method \cite{Schlessinger:1966zz, PhysRev.167.1411, Tripolt:2016cya, Chen:2018nsg, Binosi:2019ecz},
developed analytic approximations that delivered estimates for the next three.  Our prediction for ${\mathpzc q}^\pi(x;\zeta_H)$ was reconstructed from this information on the first six independent moments and the algebraically computed $x\simeq 0,1$ endpoint behaviour [Sec\,\ref{predictqpion}].

In continuum studies, the value of the hadronic scale, $\zeta_H$, has typically been a parameter; usually chosen to obtain agreement with the value of some Mellin moments determined in phenomenological analyses of data \cite{Jaffe:1980ti}.  That is not the case herein.  Instead, the value $\zeta_H=m_\alpha$,
%%\begin{equation}
%%\zeta_H=0.30\,{\rm GeV},
%%\end{equation}
Eq.\,\eqref{alphaPI}, is determined at the outset by connecting the one-loop running coupling with QCD's process-independent effective charge \cite{Binosi:2016nme, Rodriguez-Quintero:2018wma}.

Our result for ${\mathpzc q}^\pi(x;\zeta_H)$ [Eq.\,\eqref{qpizetaH}] exhibits the $x\simeq 1$ behaviour predicted by the QCD parton model, Eq.\,\eqref{PDFQCD}.  Moreover, ${\mathpzc q}^\pi(x;\zeta_H)$ is a broadened function.  As with meson distribution amplitudes, this hardening is a consequence of dynamical chiral symmetry breaking (DCSB), itself a realisation of the mechanism responsible for the emergence of mass in the Standard Model.

With the hadronic scale fixed, we used the process-independent effective charge to integrate the evolution equations and obtain ${\mathpzc q}^\pi(x;\zeta_2=2\,{\rm GeV})$ and ${\mathpzc q}^\pi(x,\zeta_5=5.2\,{\rm GeV})$ [Sec.\,\ref{secEvolution}]; and, simultaneously, predictions for the associated glue and sea quark distribution functions within the pion.  At $\zeta_2$, the scale typical of both lattice-QCD studies and phenomenological analyses of data, we determined the following momentum budget for the pion:
%%\begin{equation}
%%\langle x_{\rm valence} \rangle  = 0.48(3)\,,\;
%%\langle x_{\rm glue} \rangle  = 0.41(2)\,,\;
%%\langle x_{\rm sea} \rangle = 0.11(2)\,,
%%\end{equation}
\begin{subequations}
\begin{align}
\langle x_{\rm valence} \rangle & = 0.48(3)\,, \\
\langle x_{\rm glue} \rangle & = 0.41(2)\,, \\
\langle x_{\rm sea} \rangle & = 0.11(2)\,,
\end{align}
\end{subequations}
confirming the large gluon momentum-fraction found in earlier continuum analyses \cite{Hecht:2000xa, Chen:2016sno}.  Furthermore, our prediction for ${\mathpzc q}^\pi(x,\zeta_5)$  [Fig.\,\ref{qpizeta5}] agrees with $\pi N$ Drell-Yan data \cite{Conway:1989fs} rescaled as suggested by the complete next-to-leading-order (NLO) reanalysis in Ref.\,\cite{Aicher:2010cb}.

Of particular importance is the agreement between our parameter-free result for ${\mathpzc q}^\pi(x,\zeta_5)$ and that obtained in a recent, exploratory lQCD calculation \cite{Sufian:2019bol}.  With this confluence, two disparate treatments of the pion bound-state problem have arrived at the \emph{same} prediction for the pion's valence-quark distribution function.  This should stimulate a reconsideration of extant phenomenological analyses so that the next attempts involve a complete NLO analysis of data, including the threshold resummation effects which seem so crucial to obtaining a sound extraction of ${\mathpzc q}^\pi(x)$.  The results presented herein also support efforts to obtain new data on pion distribution functions, such as those approved at the Thomas Jefferson National Accelerator Facility \cite{Keppel:2015, Keppel:2015B, McKenney:2015xis} and identified as high priority at other facilities \cite{Petrov:2011pg, Peng:2016ebs, Peng:2017ddf, Horn:2018fqr, Denisov:2018unj}.

A worthwhile extension of the analysis described herein is the calculation of analogous kaon distribution functions.  This will enable a sophisticated reevaluation of predictions from an earlier algebraic analysis \cite{Chen:2016sno}, which indicated that the gluon content of the kaon is significantly smaller than that of the pion and identified the origin of this effect to be DCSB and its role in forming the almost-massless pion \cite{Roberts:2016vyn}.

\begin{acknowledgments}
%\centerline{\textbf{ACKNOWLEDGMENTS}}
%
%\smallskip
%
We are grateful for constructive comments from A.~Bashir, C.~Chen, M.~Chen, F.~Gao, C.~Mezrag, J.~Papavassiliou, J.~Repond, J.~Rodr{\'{\i}}guez-Quintero and J.~Segovia;
for the hospitality and support of RWTH Aachen University, III.\,Physikalisches Institut B, Aachen - Germany;
and likewise for the hospitality and support of the University of Huelva, Huelva - Spain, and the University of Pablo de Olavide, Seville - Spain, during the ``4th Workshop on Nonperturbative QCD'' (University of Pablo de Olavide, 6-9 November 2018).
Work supported by:
the Chinese Government's \emph{Thousand Talents Plan for Young Professionals};
Jiangsu Province \emph{Hundred Talents Plan for Professionals};
and Forschungszentrum J\"ulich GmbH.
\end{acknowledgments}

%%%%

%%\bibliographystyle{../../../../zProc/z10/z10KITPC/h-physrev4}
%%\bibliography{../../../../CollectedBiB}

\end{document}